\documentclass[astrosymb, resetfootnote]{aastex7}

\newcommand\chandra{{\it Chandra}}

\newcommand\xmm{{\it XMM-Newton}}
\newcommand\XMM{{\it XMM-NEWTON}}

\newcommand\swift{{\it Swift\/}}
\newcommand\nustar{{\it NuSTAR}}
\newcommand\suzaku{{\it Suzaku}}

\newcommand\gaia{{\it Gaia}}

\newcommand\fermi{{\it Fermi}}

\newcommand{\sgra}{\ensuremath{\text{Sgr\,A}^{\star} }}

\def\simgt{\lower.5ex\hbox{$\; \buildrel > \over \sim \;$}}
\def\simlt{\lower.5ex\hbox{$\; \buildrel < \over \sim \;$}}

\newcommand{\fluxcgs}{ergs~s$^{-1}$~cm$^{-2}$}
\newcommand{\lumcgs}{ergs~s$^{-1}$}

\usepackage [english]{babel}
\usepackage [autostyle, english = american]{csquotes}
\MakeOuterQuote{"}

\usepackage{comment}
\usepackage{array}
\usepackage{multirow}

\usepackage{setspace}
\usepackage{tgtermes}
\usepackage{enumitem}

\submitjournal{ApJ}

\shorttitle{Galactic Center X-ray Sources}
\shortauthors{Mandel et al.}
\begin{document}

\title{Unveiling a Concentration of Hard X-Ray Point Sources in the Nuclear Stellar Disk with XMM-Newton} 

\correspondingauthor{Shifra Mandel}
\email{ss5018@columbia.edu}

\author[0000-0002-6126-7409]{Shifra Mandel}
\altaffiliation{National Science Foundation Fellow}
\affiliation{Columbia Astrophysics Laboratory, Columbia University, New York, NY 10027, USA}
\email{ss5018@columbia.edu}

\author[0000-0002-9709-5389]{Kaya Mori} 
\affiliation{Columbia Astrophysics Laboratory, Columbia University, New York, NY 10027, USA}
\email{km211@columbia.edu}

\author[0000-0003-0293-3608]{Gabriele Ponti} 
\affiliation{INAF - Osservatorio Astronomico di Brera, Via E. Bianchi 46, Merate, 23807, Lecco, Italy}
\affiliation{Max-Planck-Institut für extraterrestrische Physik, Garching, Germany}
\affiliation{Como Lake Center for Astrophysics (CLAP), DiSAT, Università degli Studi dell’Insubria, via Valleggio 11, 
22100 Como, Italy}
\email{gabriele.ponti@inaf.it}

\author[0000-0001-9042-965X]{Samaresh Mondal} 
\affiliation{Department of Astronomy, University of Illinois, Urbana, IL 61801, USA}
\email{samaresh@illinois.edu}

\author[0000-0002-3681-145X]{Charles~J.~Hailey}
\affiliation{Columbia Astrophysics Laboratory, Columbia University, New York, NY 10027, USA}
\email{chuckh@astro.columbia.edu}

\author[0000-0003-0155-2539]{Michael M. Shara} 
\affiliation{Department of Astrophysics, American Museum of Natural History, New York, NY, USA}
\email{mshara@amnh.org}

\author[0000-0002-5082-5049]{Tong Bao} 
\affiliation{INAF - Osservatorio Astronomico di Brera, Via E. Bianchi 46, Merate, 23807, Lecco, Italy}
\email{tbao@astro.physik.uni-potsdam.de}

\author[0000-0003-2840-6152]{Heshou Zhang} 
\affiliation{INAF - Osservatorio Astronomico di Brera, Via E. Bianchi 46, Merate, 23807, Lecco, Italy}
\affiliation{Max-Planck-Institut f\"ur Physik, Boltzmannstr. 8, Garching bei M\"unchen, 85748, Germany}
\email{heshou.zhang@inaf.it}

\author[0000-0002-6753-2066]{Mark R. Morris}
\affiliation{Department of Physics and Astronomy, UCLA, Los Angeles, CA 90095, USA}
\email{morris@astro.ucla.edu}

\author[0000-0002-6089-5390]{JaeSub Hong}
\affiliation{Center for Astrophysics | Harvard \& Smithsonian, Cambridge, MA 02138, USA}
\email{jhong@cfa.harvard.edu}

\author[0000-0002-1323-5314]{Jonathan Grindlay}
\affiliation{Center for Astrophysics | Harvard \& Smithsonian, Cambridge, MA 02138, USA}
\email{jgrindlay@cfa.harvard.edu}

\author[0000-0002-3955-8189]{Nicola Locatelli} 
\affiliation{INAF - Osservatorio Astronomico di Brera, Via E. Bianchi 46, Merate, 23807, Lecco, Italy}
\email{nicola.locatelli@inaf.it }

\author[0009-0007-0196-0256]{Simone Piscitelli} 
\affiliation{INAF - Osservatorio Astronomico di Brera, Via E. Bianchi 46, Merate, 23807, Lecco, Italy}
\email{simone.piscitelli@inaf.it}

\begin{abstract}

The Galactic Center is home to the largest known concentration of X-ray sources ever detected in our Galaxy, including a sizable population of accreting compact objects.  Utilizing data from the \xmm\ Galactic Plane Heritage Survey of the central $20^{\circ}\times2^{\circ}$ of the Milky Way, we investigate the varying source populations in the inner regions of the Galaxy.  We identify for the first time a concentration of extremely hard X-ray sources within $|l|\simlt1.3^{\circ}$ and $|b|\simlt0.3^{\circ}$, a region that overlaps the nuclear stellar disk (NSD).  We determine that this hardness is intrinsic to the sources and not a manifestation of high line-of-sight column density ($N_{\rm H}$).  
Unlike the "soft" X-ray sources, this population does not trace the stellar mass distribution, indicating an overabundance compared to the underlying stellar population.  
We simulate spectra for a variety of source types to model their X-ray hardness and compare their distributions to those we observed in our survey data.  We find that a significant subset of the Galactic Center X-ray population may comprise intermediate polars (IPs) with massive white dwarfs, possibly formed through dynamical processes.  
Our study provides novel observational insights into the distribution of stellar remnants and other X-ray sources in the vicinity of our supermassive black hole, \sgra.
\end{abstract}


\section{Introduction} \label{sec:intro}

More than 44 years ago, \cite{Worrall1982} discovered what would become known as the Galactic ridge X-ray emission (GRXE) -- an apparently diffuse X-ray glow that seemed to trace the disk of the Milky Way.  In the intervening decades, many studies have attempted to discern the origin of this emission.  Some of the earliest efforts proposed that numerous unresolved faint point sources, such as cataclysmic variables (CVs) and active binaries (ABs), were responsible \citep{Worrall1983, Koyama1986a, Revnivtsev2006}.  Others speculated that the source of the emission was indeed diffuse, proposing exotic mechanisms to supply or trap a sufficiently hot plasma bound to the shallow gravitational well of the Galactic disk \citep{Koyama1986b, Yamauchi1996, Tanuma1999, Ebisawa2001, Tanaka2002}.  \cite{Revnivtsev2009} put an end to most of the speculation about the GRXE origin by showing 
that most of the observed X-ray emission can be accounted for by a population of faint point sources that were detected through deep \chandra\ observations of a low-extinction region (the "Limiting Window" [LW]). This was followed by \cite{Hong2012}, who showed that magnetic CVs (mCVs) dominated the population of faint sources in the LW -- although this conclusion was disputed by \cite{Xu2016}, who argued for a majority contribution from dwarf novae (DNe) supplemented by ABs, based on the Fe line equivalent widths (EWs) and intensity ratios of those sources.  The latter study was subsequently challenged 
by \cite{Schmitt2022}, who showed that ABs (like RS CVn systems) do not significantly contribute to the Fe emission and concluded that DNe and polars are the primary suspects.  More recently, \cite{Bao2026} has suggested that a population of symbiotic stars may comprise up to one-fifth of the GRXE emission overall, and an even larger fraction of the Fe XXV emission.  An additional excess of neutral (or low ionization) Fe I-K$\alpha$ emission (at 6.4 keV) has been attributed to the bombardment of molecular clouds by locally originating low-energy cosmic rays (LECRs) \citep{Yamauchi2016, Nobukawa2016}.

A 1989 breakthrough study with Ginga found a bright peak in the GRXE 6.7 keV (Fe XXV) emission at the location of the Galactic Center \citep{Koyama1989}, which became known as the Galactic Center X-ray emission (GCXE).  This excess, which was soon shown to extend to $|l|\sim1^{\circ}$ and $|b|\sim0.3^{\circ}$, has been extensively studied over the last three-and-a-half decades -- but its origin remains as hotly disputed as ever.  

Many studies have concluded that just as the GRXE is comprised of many faint, unresolved X-ray point sources, so too the GCXE originates from a (possibly distinct) population of faint CVs \citep{Hong2009, Heard2013, %Nobukawa2016, 
Xu2019, Anastasopoulou2023}.  Others have argued for a diffuse origin \citep{Koyama2007, Uchiyama2011, Uchiyama2013, Nishiyama2013}.  
The launch of the \suzaku\ telescope spurred a veritable tsunami of studies on the Fe line profiles of the Galactic Center and ridge. \cite{Koyama2007} pointed out that the spatial distribution of X-ray point sources resolved by \chandra\ does not follow the surface brightness profile of the diffuse GCXE, hence the origin of the GCXE must be separate from those point sources.  
Using \suzaku\ data, \cite{Yamauchi2009} resolved the neutral (Fe I-K$\alpha$; 6.4 keV), He-like (Fe XXV; 6.7 keV), and H-like (Fe XXVI; 6.9 keV) Fe-K emission lines along the Galactic plane and noted that the line flux ratios ($I_{6.4}/I_{6.7}$ and $I_{6.9}/I_{6.7}$) are higher in the Galactic Center compared to the GRXE.  Similar differences in the Fe-K line EWs and intensity ratios were used by \cite{Uchiyama2011, Uchiyama2013} to argue that the origin of the GCXE is distinct from that of the GRXE and therefore likely diffuse.   
\cite{Nishiyama2013} supported the argument for a (partially) diffuse origin of the GCXE by showing that the longitudinal Fe XXV line profile is steeper than the stellar number density (SND) indicated by the near-infrared (NIR) surface brightness along the nuclear stellar disk (NSD), a dense stellar structure of scale height $\sim200$ pc rotating around Sgr A*.  
And \cite{Nobukawa2016} argued that the combination of excess Fe I-K$\alpha$ (neutral Fe emission from cold gas), Fe XXV and Fe XXVI (highly ionized Fe from hot plasma) observed in the Galactic Center cannot be explained by any population of point sources, postulating instead a diffuse origin related to past \sgra\ flaring activity, or possibly an association with SNe in the central molecular zone (CMZ).

\cite{Anastasopoulou2023} presented an alternative hypothesis: by analyzing the diffuse Fe XXV emission from \xmm\ observations in the central $\sim20^{\circ}\times2^{\circ}$ of the Galactic disk and comparing it to a refined stellar mass distribution (SMD) model, they showed that a population of point sources with higher Fe abundances $Z_{\rm Fe}$ ($\sim1.9$ times the $Z_{\rm Fe}$ in the Galactic ridge) in the NSD region can account for nearly all of the observed Fe XXV excess.  
The existence of such a population is supported by NIR studies that suggest higher metallicity (including Fe abundances) in the Galactic Center compared to the bar/bulge \citep{Feldmeier-Krause2017, Nogueras-Lara2020, Schodel2020, Schultheis2021}.  It is also
consistent with the conclusions of \cite{Heard2013}, who used a much smaller \xmm\ survey to argue that unresolved point sources are responsible for the GCXE and ascribed the steeper X-ray surface brightness profile (compared to the SMD) to either SMD inaccuracies or a distinct X-ray source population.  Interestingly, \cite{Xu2019} found -- based on deep \chandra\ observations -- that %in 
the central 500\arcsec\ are populated by CVs hosting white dwarfs (WDs) of unusually high mass ($M_{\rm WD}$), speculating, like \cite{Heard2013}, that dynamical processes in the crowded  
Galactic Center may enhance the population of accreting compact objects relative to the remainder of the Galaxy.  Even more intriguing was their analysis showing that fainter CVs, particularly IPs with $L_X<6\times10^{31}$ \lumcgs, tend to have higher EWs and higher temperatures \citep{Xu2019}.  These findings support the conclusion by \cite{Hailey2016} that a concentration of IPs with high $M_{\rm WD}$ are responsible for the central hard X-ray emission (CHXE), discovered by \nustar\ at energies $>20$ keV in the central $\sim4\arcmin$.

Apart from -- though partly overlapping -- the GRXE and GCXE, an additional region of X-ray emission was found to extend $\approx5 \deg$ in radius from the center of the Galaxy \citep{Yamauchi1993}.  Known as the Galactic bulge X-ray emission (GBXE), it is even more diffuse than the GRXE and though it shares some similarities with the latter \citep{Revnivtsev2006}, differences have been noted. In particular, \cite{Nobukawa2016} argued for a majority population of non-magnetic CVs, disputing both \cite{Revnivtsev2009} and \cite{Hong2012}, who postulated more significant contributions from ABs and mCVs, respectively.  The hypothesis of \cite{Nobukawa2016} was supported by \cite{Perez2019}, who found that the hard X-ray continuum in the inner $1^{\circ}-3^{\circ}$ of the Galactic bulge resembles a thermal plasma with a temperature of $kT\approx8$ keV, consistent with (quiescent) DNe.  Unlike \cite{Yuasa2012}, \cite{Perez2019} found no evidence for a large population of intermediate polars (IPs) in the bulge with \nustar; this discrepancy is ascribed to the differences in the regions probed by these respective studies, with \cite{Yuasa2012} extending into the GCXE region and down to $r<1^{\circ}$, thought to be dominated by IPs with massive WDs \citep{Hailey2016, Xu2019}.

In recent years, the \xmm\ telescope conducted the unprecedented Galactic Plane Heritage Survey, scanning the central $\sim20^{\circ}\times2^{\circ}$ of the Galactic plane.  This survey, spanning an area that is approximately an order of magnitude larger than previous studies of X-ray point sources in the Galactic Center region (e.g., \cite{Muno2009}), offers a unique opportunity to uncover the X-ray populations in the inner regions of the Galaxy and answer some of the lingering questions about the origins and composition of the Galactic diffuse X-ray emission, including the GRXE and the GCXE.  
While previous smaller studies conducted in different regions with different instruments often yielded inconsistent results, this survey for the first time allows us to analyze the hard X-ray data over an extensive area in a consistent manner.  
Here we present an expansive study of the X-ray point sources within the Heritage Survey region ($|l|<10^{\circ}$ and $|b|<1^{\circ}$).  
We systematically explore the distributions of sources and their X-ray properties, and how they vary in different regions.  
In Section \ref{sec:obs}, we describe the \xmm\ survey and data processing.  Section \ref{sec:hr} details our analysis of the point sources within the survey region.  Section \ref{sec:sims} is dedicated to the simulations of several types of X-ray sources using spectral models.  In Section \ref{sec:distr}, we analyze the spatial and luminosity distributions of our sample.  We discuss the results of our analysis -- and their implications -- in Section \ref{sec:discuss}, and summarize our findings in Section \ref{sec:conclude}. \\

\setlength{\tabcolsep}{1.9pt}
\begin{table}[]
    \centering
    \caption{List of structures and regions of interest in and around the Galactic Center.  Note that at the Galactic Center, $1$ pc $\approx25\arcsec$.  "Diffuse" X-ray emission refers to apparently extended features that may originate from large numbers of unresolved point sources.}
    \label{tab:abbr}
    \begin{tabular}{l | l | l | l}
    \hline\hline
    Name
    & Full name & Extent/scale height & Description \\
    \hline
        GRXE & Galactic ridge X-ray emission & $|l|\sim100^\circ$, $|b|\sim1^\circ$ & "diffuse" X-ray emission along the Galactic disk \\
        GBXE & Galactic bulge X-ray emission & $r\sim5^{\circ}$ from Sgr A* & "diffuse" X-ray emission associated with the Galactic bulge \\
        GCXE & Galactic Center X-ray emission & $|l|\simlt1^\circ$, $|b|\sim0.3^\circ$ & "diffuse" Fe XXV X-ray emission \\
        NSD & nuclear stellar disk & $|l|\simlt1.3^\circ$, $|b|\sim0.3^\circ$ & dense stellar structure rotating around Sgr A* \\
        CMZ & central molecular zone & $-0.7<l<1.7$, $|b|<0.2$& giant complex of dense molecular clouds \\ 
        CHXE & central hard X-ray emission & $r\sim4\arcmin$ ($\sim10$ pc) 
        & "diffuse" hard ($kT>35$ keV) X-rays uncovered by \nustar\ \\
        NSC & nuclear star cluster & $r\sim2\arcmin$ ($\sim5\pm1$ pc) & dense massive stellar cluster surrounding Sgr A* \\ [0.3ex]
        \hline
    \end{tabular}
\end{table}

\section{Observations and Data Selection} \label{sec:obs}

\begin{figure}
\centering
\includegraphics[width=\textwidth]{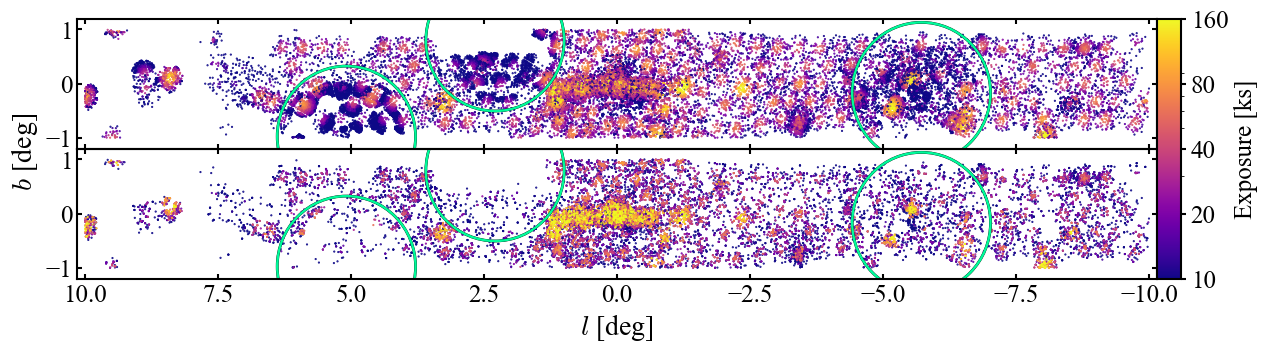}
\caption{Top: all \xmm\ sources listed in DR14 that fall within the Heritage Survey region covering the central $20^{\circ}\times2^{\circ}$ of the Galactic disc.  Bottom: the final sample used in all analyses, after filters described in Section \ref{sec:obs} were applied.  Exposure is shown in color.  Green circles mark regions contaminated by stray light background.   
}
\label{fig:exp}
\end{figure}

The \xmm\ Heritage Survey of the Galactic plane (proposal ID 088601, PI: G. Ponti) was conducted between March 13, 2021 and April 6, 2024 using the EPIC MOS and PN instruments.  It covered the region spanning $|l|<10^{\circ}$ 
in Galactic longitude and $|b|<1^{\circ}$, latitude.  Exposures ranged between $\sim40$ ks for the innermost region and $\sim20$ ks for the farther disk regions.  
In addition to the new scans of the Galactic disk, we also utilized archived data from earlier \xmm\ observations of the Galactic Center, some of which were considerably deeper.  These data are particularly useful in the innermost arcminutes of the Galaxy, where a combination of increased source density, background X-ray emission, and absorption hampers source detection.    

The \xmm\ Serendipitous Source Catalog (XMMSSC) Fourteenth Data Release (4XMM-DR14), published in July 2024, contains all \xmm\ observations completed through December 2023, including most of the Heritage Survey of the inner Galactic plane.  Crucially, the DR14 data \citep{Webb2020} were visually screened for spurious sources, which were a significant source of contamination in the crowded Galactic Center region covered by the Heritage Survey.  Therefore, using the DR14 catalog data allowed us to select a "clean" sample of sources throughout the survey region.  
Figure \ref{fig:exp} (top) shows the DR14 catalog sources within the regions covered by the \xmm\ Heritage Survey of the inner Galactic plane along with their corresponding exposures.

We removed from our analysis all DR14 catalog entries flagged as potentially spurious or as having unreliable parameters, as well as sources flagged as "extended" or "confused".  We also remove potential duplicate entries of the same source in the catalog.   
Our final tally included a total of 14753 individual unique sources, 14512 of which were detected in the $2-12$ keV band, within $|l|<10^{\circ}$ and $|b|<1^{\circ}$.  
The latter sample was utilized for all following data analysis.  Further details on the catalog source selection, elimination and consolidation can be found in Appendix \ref{sec:AA}.

Figure \ref{fig:exp} (bottom panel) shows the sources that remained after the aforementioned filters were applied. Table \ref{tab:filters} lists the filters that were applied to the \xmm\ dataset throughout different stages of the analysis -- including the processes described in the following sections -- and the resulting sample sizes.  We note that the selection process described above and in Appendix \ref{sec:AA}, though ensuring the "cleanest" sample, disproportionately affects the densest innermost regions of the survey; see Appendix \ref{subsec:AAb} for more.    \\

\renewcommand{\arraystretch}{1.1}
\setlength{\tabcolsep}{16pt}
\begin{table}[]
    \centering
    \begin{tabular}{l r l}
    \hline\hline
    Filter description & \multicolumn{1}{c}{Remaining sources:} & \S  \\[0.2em]
  & PN\,/\,MOS \qquad\qquad %\hfill 
  All & \\[0.2em]
        \hline
        None; all 4XMM-DR14 sources within $|l|<10^{\circ}$ and $|b|<1^{\circ}$ &  
        48073 & \ref{sec:obs} \\
        \multirow{2}{10.5cm}{Removal of extended sources and those flagged %as 
        "possibly spurious/located %in an area 
        where spurious detection may occur%", or both% (\textsc{SUM\_FLAG}=$2-4$)
        %; sources flagged as "
        /source parameters may be affected" %(\textsc{SUM\_FLAG}>0)
        } 
        & \multirow{2}{0.9cm}{25711} & \multirow{2}{0.75cm}{\ref{sec:obs}} \\
        & & \\
        Removal of potential duplicate sources (with nearest neighbor $<6\arcsec$ offset) & 14753   & \ref{sec:AA} \\
        Removal of sources not detected in the $2-12$ keV band & 14512   & \ref{sec:obs} \\
        Removal of sources with $\texorpdfstring{\sigma\textsubscript{HR}}{}>0.35$ & 6495\,/\,7911 \hfill 9933 & \ref{sec:hr} \\
        Removal of foreground sources & 5983\,/\,7404 \hfill 9190 & \ref{subsec:AAa}  \\
        \hline
        Selection of sources with count rate above completeness threshold & \multicolumn{1}{l}{4727\,/\,4812}  & \ref{subsec:test_hr} \\
%        \hspace{4.0cm} EPIC-PN  \qquad ($2\times10^{-3}$ cts s$^{-1}$) \hfill  & \multicolumn{1}{l}{5042} &  \\
%        \hspace{4.0cm} EPIC-MOS \,\quad ($1\times10^{-3}$ cts s$^{-1}$) \hfill  & \multicolumn{1}{l}{5078} & \\
        Selection of sources with exposures in uniform range %(PN: $22-53$ ks, MOS: $60-138$ ks) 
        & \multicolumn{1}{l}{1033\,/\,1010} & \ref{subsec:test_hr} \\
%        \hspace{4.0cm} EPIC-PN \qquad ($22<\rm exposure<53$ ks) \hfill   & \multicolumn{1}{l}{1105} &  \\
%        \hspace{4.0cm} EPIC-MOS \quad ($60<\rm exposure<138$ ks) \hfill   & \multicolumn{1}{l}{1089} & \\
        Selection of sources within $|b| < 0.2^{\circ}$ (for HR distribution along $l$) & \multicolumn{1}{l}{2291\,/\,2908} & \ref{subsec:trans} \\
        Selection of sources within $|l| < 1^{\circ}$ (for HR distribution along $b$) & \multicolumn{1}{l}{2415\,/\,2996} & \ref{subsec:trans} \\
        Selection of sources with $f_X>5\times10^{-14}$ \fluxcgs & \multicolumn{1}{l}{2274\,/\,2783} & \ref{sec:shdist}  \\
        Selection of sources with $f_X>10^{-13}$ \fluxcgs & \multicolumn{1}{l}{1086\,/\,1418} & \ref{subsec:mCVs} \\
        \hline
    \end{tabular}
    \caption{List of filters applied to the dataset in the following sections, the corresponding number of sources remaining in the sample after the filter was applied, and the section in which the process is described.  The first six filters were applied consecutively and to all subsequent samples.  }
    \label{tab:filters}
\end{table}

\section{Hardness Ratio Analysis} \label{sec:hr}

Most of the sources in our sample do not have sufficient counts for statistically sound spectral analysis.  We therefore utilized \emph{hardness ratio} (HR) as a proxy for spectral classification.  Hardness ratio is defined as 
\begin{equation}\label{eq:HR}
    \textrm{HR}\equiv\frac{H-S}{H+S}
\end{equation}
where $H$ and $S$ denote the "hard" and "soft" count rates%\footnote{We utilize count rates to calculate HR because the former is vignetting-corrected, unlike counts alone.  We do, however, use the number of counts detected to calculate the uncertainty $\texorpdfstring{\sigma\textsubscript{HR}}{}$.}
, respectively.  We define the energy bands for $H$ and $S$ as $4.5-12.0$ keV and $2.0-4.5$ keV, respectively (bands 4 and 5 in the DR14 catalog).  These energy bands are helpful for distinguishing between hard and soft sources in the Galactic Center region, where high $N_{\rm H}$ ($>10^{22}$ cm$^{-2}$ \citep{Zhu2018, Battersby2024}) results in the absorption of X-rays at energies $\simlt2$ keV.  

We then evaluated the uncertainties related to the HR values for each instrument.  The uncertainty $\sigma$ is defined by combining in quadrature the deviations in HR resulting from shifting the values of $H$ and $S$ by their respective uncertainties, following equation 1.31 in \cite{Lyons1991}: 
\begin{equation}\label{eq:HRerr}
    \texorpdfstring{\sigma\textsubscript{HR}}{}\equiv\sqrt{({\rm{HR_0}-HR}_H)^2+({\rm HR_0-HR}_S)^2}
\end{equation}
where $\rm HR_0=$$f(H,S)$ as defined by equation \ref{eq:HR}; $\rm HR$$_H=f(H+\sigma_H,S)$; and $\rm HR$$_S=f(H,S+\sigma_S)$, where $\sigma_H$ and $\sigma_S$ represent the uncertainties in the hard and soft X-ray counts, respectively.

To further maximize the statistical robustness of our sample, we proceeded to remove sources with $\texorpdfstring{\sigma\textsubscript{HR}}{}>0.35$; this threshold was refined based on the difference between the average HR in the two regions described below.  Our sample 
then contained 6495/7911 sources -- detected by PN/MOS -- with relatively well constrained HR.  Of those sources, $>70\%$ (4675 sources) are present in both the PN and MOS sample, while the remainder were either not detected or had uncertainties $\texorpdfstring{\sigma\textsubscript{HR}}{} >0.35$ for one of the instruments.

The effective areas of the \xmm/EPIC PN and MOS instruments differ, with the individual MOS cameras having much lower effective areas in general –– and a lower ratio of hard-to-soft band effective area in particular -- compared to PN.  As a result, hardness ratios derived from the PN and MOS are not equivalent.  We therefore calculated HR for PN and MOS data separately.  In order to analyze the HR values calculated from the PN and MOS data on an equivalent basis, we apply a correction factor to the MOS count rates in the hard ($4.5-12$ keV) band, which accounts for the lower hard-to-soft ratio of effective areas in MOS compared to the PN instrument.  Details on the determination and testing of this correction factor are provided in Appendix \ref{sec:AAa}. These altered HR values are used for the MOS data in all analyses where PN and MOS data are jointly evaluated. 

Finally, we define a "foreground" hardness ratio HR$_{\rm fgd}$ with a soft band of $0.5-1$ keV and a hard band of $2-12$ keV, to help distinguish unabsorbed foreground sources from those near the Galactic Center.  We remove from the analysis sample any sources with HR$_{\rm fgd}<-0.1$ (PN; HR$_{\rm fgd}<0$ for MOS), which are likely located within $\sim3.5$ kpc.  Details on the HR$_{\rm fgd}$ distributions and threshold determination can be found in Appendix \ref{subsec:AAa}.  

Figure \ref{fig:hr1} shows the results of our hardness ratio analysis on the final sample.  While most of the sources scattered throughout the Galactic plane have relatively "soft" colors, with average values of $\simeq0.1$ for PN and $\simeq-0.3$ for MOS, many of the remaining sources in the innermost disk region ($|l|<1.3^{\circ}$ and $|b|<0.3^{\circ}$) are distinctly harder, with average HR $\simeq0.5$ for PN and $\simeq0.3$ for MOS.  The region corresponding to this concentration of hard X-ray sources is consistent with the nuclear stellar disk (NSD).  Kolmogorov–Smirnov (K–S) tests indicate that the HR distributions inside and outside the NSD are different well above the $20\sigma$ level for both PN and MOS data.

\begin{figure}
\centering
\includegraphics[width=\textwidth]{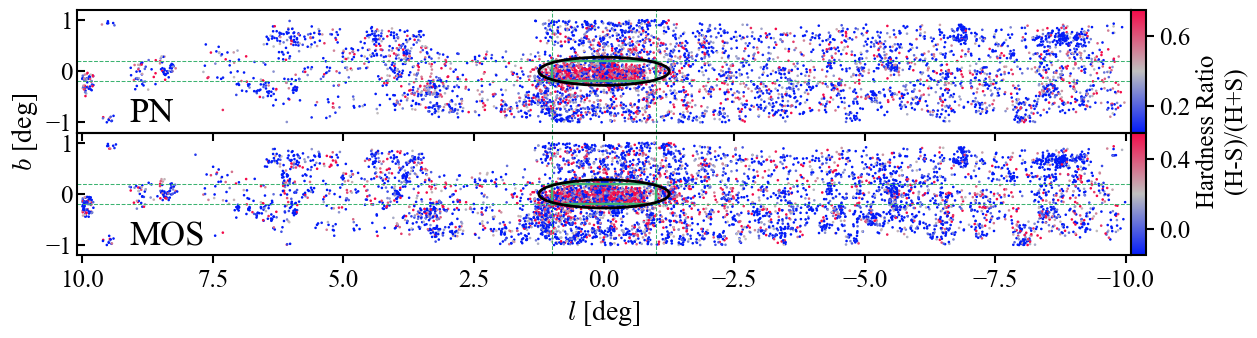}
\caption{Hardness ratio of X-ray point sources throughout the survey region from PN (top) and MOS (bottom).  $H$ and $S$ represent the $4.5-12.0$ keV and $2.0-4.5$ keV counts, respectively.  Blue indicates "soft" sources; red corresponds to "hard" sources.  The central region corresponding to the NSD (black ellipse) holds a larger concentration of "harder" (red) X-ray sources.  Green lines demarcate the regions used in Section 
\ref{subsec:trans}.  See Section \ref{sec:hr} for more.}
\label{fig:hr1}
\end{figure}

\subsection{HR tests} \label{subsec:test_hr}

Exposures for individual observations in this survey varied widely (see Figure \ref{fig:ctrt_all} in Appendix \ref{sec:AB}).  Since longer exposures often yield the detection of fainter sources due to increased sensitivity, this could potentially lead to biases in the observed HR distributions for regions with differing exposures.  Specifically, the central $\sim1^{\circ}$ of the Galaxy was subject to much higher average exposures compared to the remainder of the survey (Figure \ref{fig:exp}).  This might render the survey more sensitive towards faint sources in the central degree, despite the much higher background emission and source confusion within that region.
To check whether the X-ray source population observed in the central region is \emph{intrinsically} harder than the population in the outer disk, we conducted several tests that were designed to minimize observational biases due to differences in exposures and sensitivity between the two regions.  For the analysis described in Section \ref{subsec:test_hr}, we defined the NSD region as $|l|<1.3^{\circ}$ and $|b|<0.3^{\circ}$.

\begin{figure}
\centering
\includegraphics[width=\textwidth]{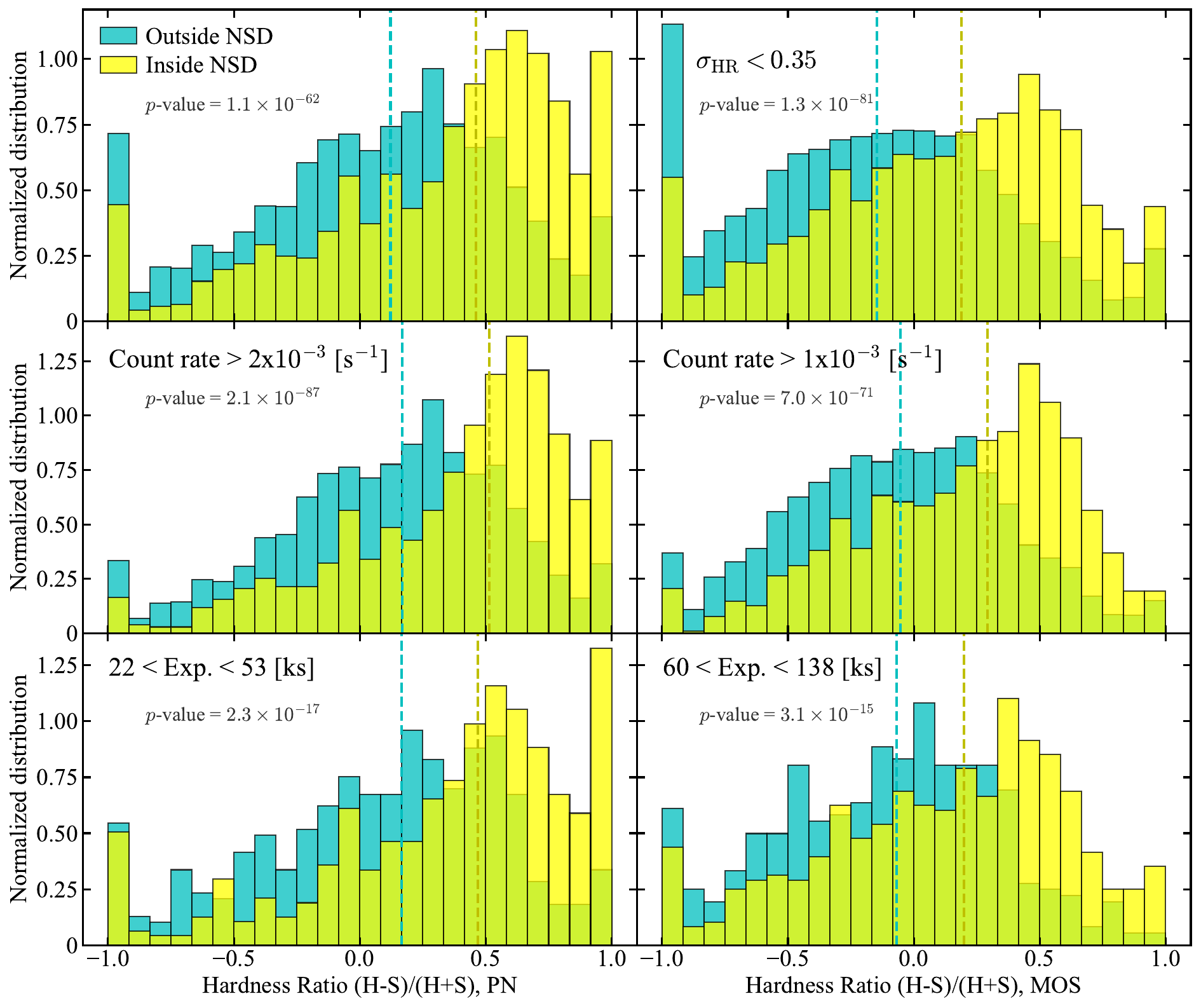}
\caption{Hardness ratio (HR) distributions of \xmm\ sources outside the NSD (cyan) and within $|l|<1.3^{\circ}$ and $|b|<0.3^{\circ}$ (yellow), normalized such that each distribution integrates to 1, for PN (left) and MOS (right).  Dashed lines mark the median values of the corresponding distributions.  
Top panel: all sources that meet the criteria described in Sections \ref{sec:obs} with $\texorpdfstring{\sigma\textsubscript{HR}}{}<0.35$.  Middle panel: subset of sources shown in top panel, with count rates above the given threshold.  Bottom panel: subset of sources shown in top panel, with exposures in the specified range.  $p-$values from K-S tests are listed at the 99.9\% level; see Table \ref{tab:filters} for sample sizes.  The differences in the HR distributions inside and outside the NSD appear to be independent of observational biases.  }
\label{fig:hr3}
\end{figure}

For our first test, we selected only sources with mean 2-12 keV count rates exceeding $10^{-3}$ s$^{-1}$ for MOS and $2\times10^{-3}$ s$^{-1}$ for PN.  These lower limits were fixed in order to filter away most sources fainter than the completeness limit of our survey (see Figure \ref{fig:ctrt_dist} in Appendix \ref{sec:AB} for the count rate distributions).  Approximately $5000$ sources, about one third of which are located inside the NSD region, exceed the minimum count rate threshold.  As Figure \ref{fig:hr3} (center row) shows, the resulting HR distributions within and outside the NSD retain their differences, with "inner" sources showing significantly 
harder X-ray distributions.

For the second test, we selected only sources with exposures ranging between $22-53$ ks (for PN) and $60-138$ ks (MOS1 and MOS2, combined).  The range was selected to impose roughly uniform sensitivity across the survey area -- i.e., inside and outside the NSD  (cf. Figure \ref{fig:ctrt_dist}, right) -- while ensuring a sufficiently robust sample size for each of the two regions/instruments ($\simgt500$ sources for each sample).  This filter reduced the potential biases introduced by the 
very long ($>150$ ks) exposures in the innermost region of the Galaxy.  Figure \ref{fig:hr3} (bottom) demonstrates that even with this filter applied, the differences in the HR distributions inside and outside the NSD persist. 

We performed K–S tests for each of the test datasets described above.  In all cases, the K–S statistics indicated that the HR distributions inside and outside the NSD region are distinct (at a level $>8\sigma$.) \\

\subsection{Where does the transition occur?} \label{subsec:trans}

In order to constrain the location of the borders that separate the region hosting a concentration of harder X-ray sources from the rest of the Galactic disk, we evaluated the average HR values -- and how they change -- over the range of Galactic latitudes and longitudes in our survey.  

\begin{figure}
\centering
\includegraphics[width=0.8\textwidth]{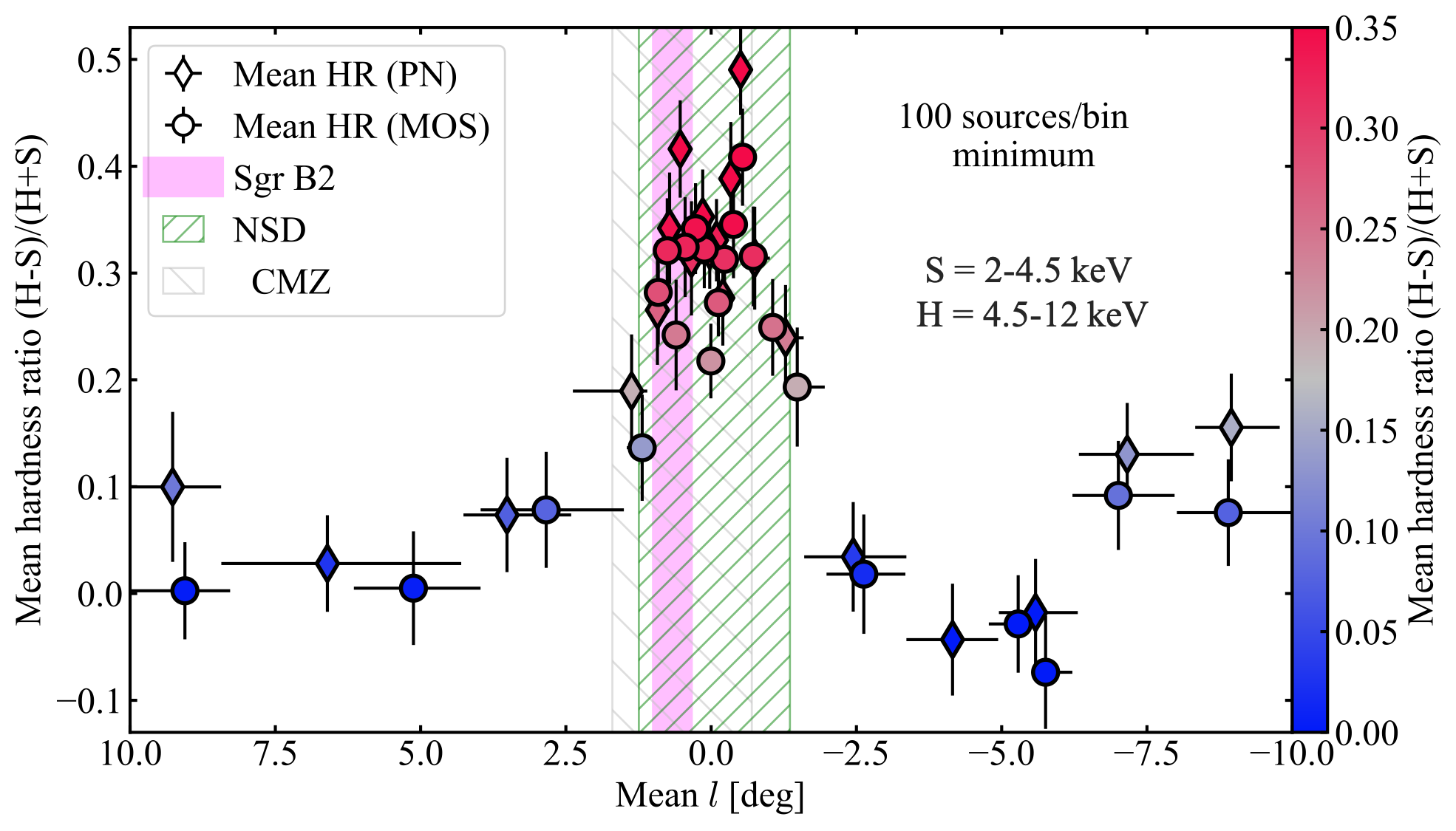}
\caption{Hardness ratio (HR) as a function of Galactic longitude.  
Diamonds and circles represent the mean values for PN and MOS, respectively, with colors corresponding to the mean HR for the bin.  Vertical error bars show the associated $1\sigma$ uncertainties for the mean HR values; horizontal error bars denote the range of longitudes spanned by the sources in each bin.  
The footprint of the NSD (green hatch), giant molecular cloud Sgr B2 (magenta), and the extended CMZ (silver hatch) are marked as well.  
Average hardness appears much higher 
within the NSD footprint.  } 
\label{fig:degbin}
\end{figure}

To determine the longitudinal points of transition, we selected X-ray sources within a narrow range of Galactic latitudes, $|b| < 0.2^{\circ}$ (sources that fall between the horizontal green dashed lines in Figure \ref{fig:hr1}).  This range was selected conservatively to exclude much of the periphery of the NSD.  2291/2908 sources fall within that range from the PN/MOS datasets, respectively (and also surpass the selection criteria described in Sections \ref{sec:obs} and \ref{sec:hr}, i.e. they are considered not likely to be spurious and their hardness ratios are well constrained).  We sorted the sources into bins of minimum width $\sim0.1^{\circ}$, containing at least 100 sources each, and calculated the mean HR for each bin.  As Figure \ref{fig:degbin} shows, the mean HR values increase at $|l|<1.3^{\circ}$. Outside of this region there is a distinctive drop in average hardness.

Significantly, the concentration of harder sources does \emph{not} appear to be aligned with the high-extinction CMZ.  Curiously, the easternmost part of the CMZ ($1.2^{\circ}\simgt l \simgt 1.7^{\circ}$) seems to host X-ray sources that are not significantly harder on average than those outside the NSD.  By contrast, the hard X-ray population in the western part of the NSD extends beyond the CMZ region.  This finding is noteworthy because it suggests that the harder sources we observe are not an artifact of absorption within the high-density CMZ.  The hard concentration does appear to follow the footprint of the NSD, further supporting an association.

\begin{figure}
\centering
\includegraphics[width=0.8\textwidth]{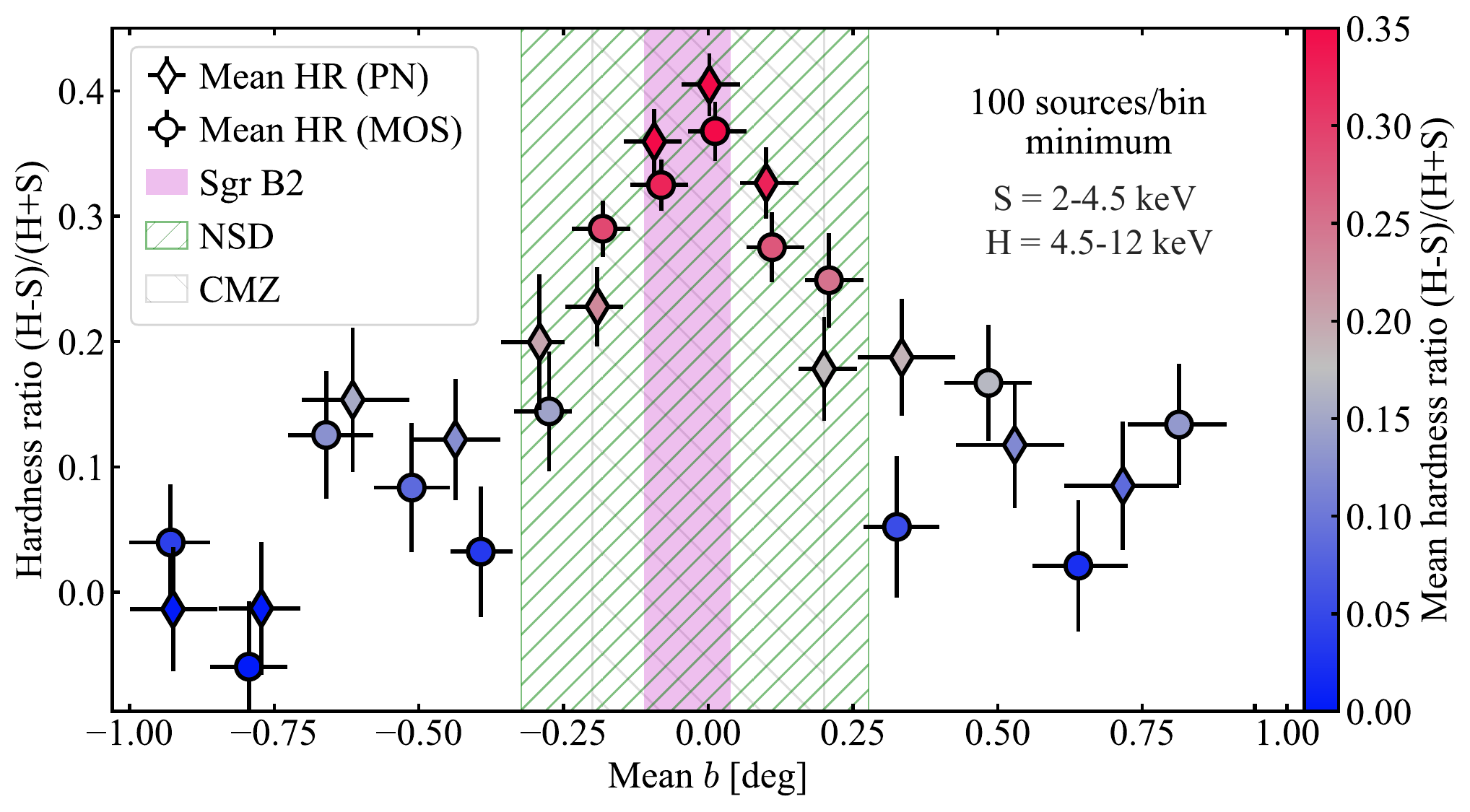}
\caption{Hardness ratio (HR) as a function of Galactic latitude.  
Markers are the same as for Figure \ref{fig:degbin}.
See Section \ref{subsec:trans} for more.}
\label{fig:degbin_b}
\end{figure}

\smallskip
We repeat this analysis for sources within $|l| < 1^{\circ}$ (vertical green dashed lines in Figure \ref{fig:hr1}) to find the transition latitude.  This sample consists of 2415/2996 sources from the PN/MOS datasets.  As Figure \ref{fig:degbin_b} shows, the point of transition to a harder source population along the innermost latitudes appears less pronounced.  
Hardness increases inside $|b|\sim0.25^{\circ}$, once again consistent with the NSD footprint.  

%\smallskip

\subsection{Are the harder sources an artifact of 
high $N_{\rm H}$?} \label{subsec:nhsims}

The Galactic Center region known as the CMZ ($-0.7^{\circ}<l<1.7^{\circ}$, $|b|<0.2^{\circ}$) is home to a large number of dense molecular clouds and exhibits high levels of extinction \citep{Nogueras2021, Battersby2024}.  When interpreting X-ray data from a limited bandpass, it is sometimes impossible to determine whether a source is intrinsically hard or just heavily absorbed in the soft band -- unless the absorption can be independently constrained.  While $N_{\rm H}$ can be degenerate with key spectral parameters, the degeneracy can be broken if the line-of-sight absorption is known.  We therefore utilize the detailed 3-D CMZ studies conducted by \cite{Battersby2024} to analyze how the average absorption varies over our survey region, and whether the observed variations in $N_{\rm H}$ are correlated to differences in the mean hardness ratios we measured in Section \ref{sec:hr}.  Such a correlation would indicate that the observed increase in hardness may be due to higher extinction in the innermost disk region, rather than an intrinsically harder source population.

We extracted the $N_{\rm H}$ data within $|l|\simlt10^{\circ}$ and $|b|<0.2^{\circ}$ (corresponding to the region used to evaluate the X-ray hardness distribution over Galactic longitudes), and grouped the data into bins of width $l=0.1^{\circ}$, height $b=0.4^{\circ}$ (i.e. $b\pm0.2^{\circ}$).  
For each bin, we calculated the mean and the associated uncertainty.  
Since the 3-D CMZ survey measures $N_{\rm H_2}$, we also add the line-of-sight H I column density.  
H I is notoriously difficult to constrain, as it is self-absorbing and can present in a variety of molecular configurations.  
Here we assume that the density of atomic H I is equal to molecular H$_2$ along the Galactic disk \citep{Li1983}.  This is a conservative estimate meant to account for the various forms of H contributing to the total column density; though it is an approximation, reasonable variations on the scale expected would not significantly alter our results.   However, in the CMZ, most of the "excess" H$_2$ (above the level of the surrounding Galactic disk) is contained within extremely dense molecular clouds ($N_{\rm H_2}>10^{22}$ cm$^{-2}$); the high density shields most of the molecules from dissociation, allowing H$_2$ to strongly dominate \citep{Skalidis2024}.  Indeed, the H I survey by \cite{Pohl2022} shows a relatively low fraction of H I towards the Galactic Center at a distance of $\sim8$ kpc compared to the foreground.  Therefore, we determine the mean \emph{foreground} column density $N_{\rm H\,I}$ by averaging the molecular column density $N_{\rm H_2}$ at $6^{\circ}<|l|<10^{\circ}$, where the aforementioned 1:1 correlation between $n_{\rm H I}$ and $n_{\rm H_2}$ is applied.  We then add this mean foreground $N_{\rm H\,I}$ uniformly to the 3-D CMZ-derived $N_{\rm H_2}$ to obtain a combined column density, $N_{\rm H_{tot}}=N_{\rm H\,I}+2N_{\rm H_2}$.

\begin{figure}
\centering
\includegraphics[width=.8\textwidth,trim=0.2cm 0.5cm 0.2cm 0.5cm]{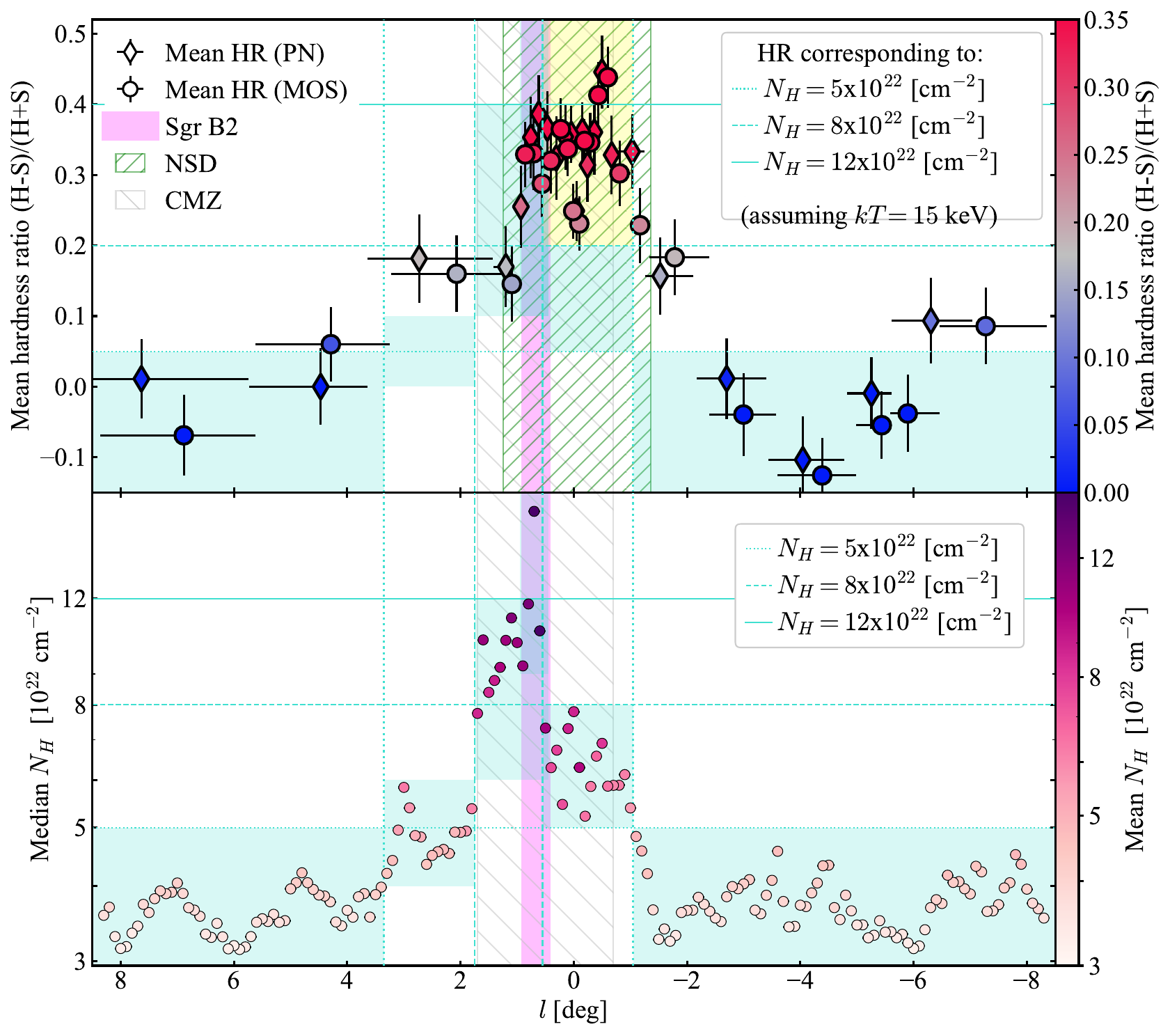}
\caption{Top:  Same as Figure \ref{fig:degbin}, with the addition of HR thresholds corresponding to mean $N_{\rm H}$ (horizontal turquoise lines).  Vertical lines denote the regions where the $N_{\rm H}$ thresholds (marked by turquoise blocks in lower panel) are surpassed.  Shaded turquoise regions mark the parameter space that is expected to be occupied, assuming a thermal \textit{apec} model with $kT=15$ keV and corresponding $N_{\rm H}$.  In the yellow shaded region, the hardening in the mean HR is inconsistent with the $N_{\rm H}$ observed at these Galactic longitudes.  Bottom: mean $N_{\rm H}$ as a function of $l$, as measured along $|b|<0.2^{\circ}$ in bins of width $0.1^{\circ}$ in longitude.  Turquoise blocks mark the range of $N_{\rm H}$ used to calculate the HR marked out by the corresponding blocks in the top panel.   
See Section \ref{subsec:nhsims} for more.}
\label{fig:hrvnh}
\end{figure}

Figure \ref{fig:hrvnh} (bottom) shows the mean $N_{\rm H}$ as a function of Galactic longitude $l$.  (Note that the $1\sigma$ uncertainties on the mean $N_{\rm H}$ are also plotted, but the error bars are smaller than the markers and therefore are not visible.)  We find that the trend in $N_{\rm H}$ does not generally correspond to the changes in observed average hardness of the X-ray sources in our survey.  This is especially obvious in the region spanning $-1.0^{\circ}<l<0.5^{\circ}$, where the average X-ray hardness is at its peak while $N_{\rm H}$ values are intermediate.  Rather, the mean $N_{\rm H}$ appears to increase along the CMZ and peaks at Sgr B2.

To further test whether the observed differences in HR can be attributed to variations in $N_{\rm H}$, we conducted simulations -- assuming a range of $N_{\rm H}$ values -- for several different phenomenological models.  Chemical abundances were adopted from \cite{Wilms2000}.  We set several different thresholds for $N_{\rm H}$, corresponding to intervals where $N_{\rm H}$ increases (turquoise blocks in Figure \ref{fig:hrvnh}, bottom).  
We simulated spectra for each model for the range of $N_{\rm H}$ thresholds and computed the resulting HR.  We find that the differences in $N_{\rm H}$ observed throughout the survey region cannot account for the observed differences in HR \emph{under the assumption that the underlying source populations are identical inside and outside the NSD}.  This holds for every spectral model we simulated.  Figure \ref{fig:hrvnh} (top) shows the observed mean HR (diamond markers) as a function of $l$.  The HR thresholds corresponding to the aforementioned $N_{\rm H}$ thresholds (calculated from simulated absorbed APEC spectra with $kT=15$ keV, the plasma temperature measured for the GRXE by \cite{Yuasa2012}) are marked in turquoise.  Matching vertical lines denote the longitude $l$ at which each corresponding $N_{\rm H}$ threshold is surpassed.  Shaded turquoise regions mark the range of HR values expected based on the corresponding mean $N_{\rm H}$, assuming a thermal \textit{apec} model with $kT=15$ keV.  In the shaded yellow region along $-1.05^{\circ}<l<0.55^{\circ}$, the observed mean HR ($\simgt0.3$) significantly exceeds the projected HR ($<0.2$) for the corresponding mean $N_{\rm H}$ ($5\times10^{22}<N_{\rm H}<8\times10^{22}$ cm$^{-2}$).  This indicates an intrinsically harder population of X-ray sources within that region compared to the Galactic disk, independent of absorption effects.  \\

\section{Simulations of point sources}\label{sec:sims}

As explained in Section \ref{sec:hr}, we use HR as a proxy for spectral fitting of the point sources in our survey, as most of these sources lack sufficient counts for robust spectral analysis.  In order to link the measured HR values to physically meaningful characteristics, we conducted extensive simulations with a range of phenomenological models that correspond to different types of X-ray sources.  The HR values derived from these simulations can then help us classify the observed sources with corresponding HR.  We chose to simulate \xmm/EPIC-PN data, as PN has a higher effective area compared to MOS.

\subsection{Model selection and parameters}\label{sec:sim_models}

We used the \texttt{fakeit} command in \texttt{XSPEC} to generate simulated spectra, applying the \xmm/EPIC-PN RMF and ARF files.  Tables \ref{tab:apecsims} and \ref{tab:powsims} list the simulation models and their parameters.  For non-thermal sources, we utilized an absorbed powerlaw model \texttt{tbabs*powerlaw} with photon index $\Gamma$ ranging from $0-3$ in increments of 0.5.  This range is consistent with sources like quiescent LMXBs \citep{Hailey2018, Mori2021}, MSPs \citep{Lee2023, Sullivan2025}, and wind-wind collision systems \citep{Mori2017, Hamaguchi2018, Archer2020}, among others.  We simulated optically thin thermal sources using an absorbed \texttt{APEC} model \citep{Smith2001}, with temperature $kT$ spanning $1-36$ keV.  At the low end of the temperature range, these models are consistent with non-magnetic CVs and active binaries \citep{Koyama2024}; magnetic CVs tend to be much harder, with $kT$ for polars typically ranging between $5-15$ keV \citep{Xu2016, Filor2025} and intermediate polars (IPs) having mean $kT=34$ keV \citep{Xu2016}.  Each model was tested with a range of absorption $N_{\rm H}$ between $(1-30)\times10^{22}$ cm$^{-2}$, corresponding to the thresholds described in Section \ref{subsec:nhsims}.  Chemical abundances were assumed based on \cite{Wilms2000}.

We note that the Fe-K$\alpha$ emission lines, with energies between $6-7$ keV, fall in the "hard" band of our HR.  Previous studies have suggested a source population with higher metallicities or Fe abundances in the central regions of the Galaxy (the NSD and the NSC) compared to the Galactic bar/bulge \citep{Feldmeier-Krause2017, Nogueras-Lara2020, Schodel2020, Schultheis2021, Anastasopoulou2023}.  To understand how Fe abundance $Z_{\rm Fe}$ affects the derived HR values, we simulated the thermal \texttt{APEC} models using a range of $Z_{\rm Fe}/Z_{\odot}$ between $0.5-2.0$, each one with the same range of $N_{\rm H}$ and $kT$ such that each set of [$N_{\rm H}$, $kT$] was simulated with the range $Z_{\rm Fe}=[0.5, 1, 2]\times Z_{\odot}$.

Altogether there were 274 models with unique combinations of parameters, listed in Tables \ref{tab:apecsims} and \ref{tab:powsims}.  For each of the 274 models, we generated a set of 1000 spectra, each one randomized to account for statistical fluctuations.  All simulations assumed an exposure of 20 ks, consistent with the most common value for the survey (Figure \ref{fig:ctrt_all}).  
For each simulated spectrum, we then extracted the counts in the $2.0-4.5$ keV and $4.5-12.0$ keV range, corresponding to our "soft" and "hard" energy bands, respectively.  We derived the resulting HR and corresponding uncertainty $\texorpdfstring{\sigma\textsubscript{HR}}{}$ for all 274000 spectra. 

\begin{figure}
    \centering
    \includegraphics[width=0.95\linewidth]{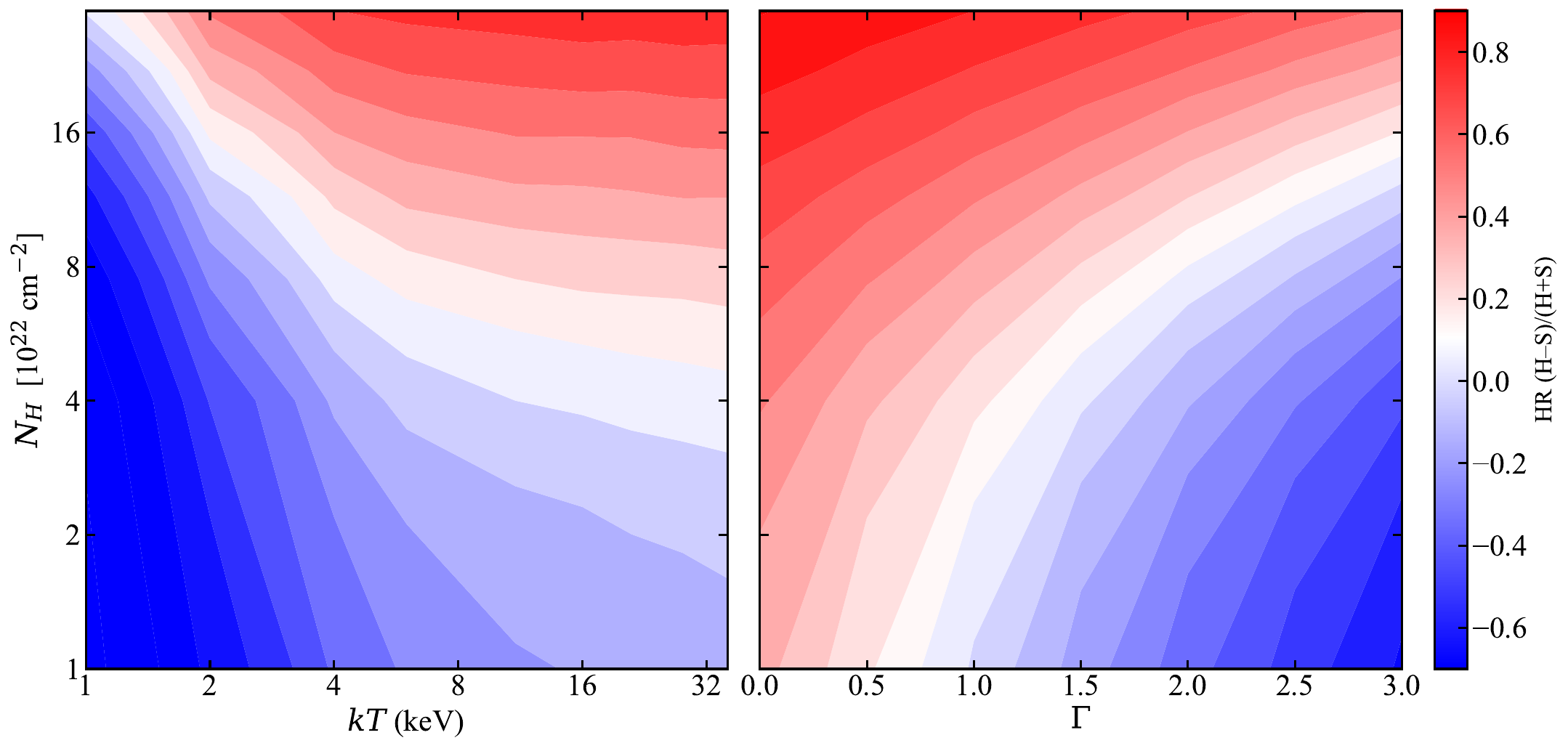}
    \caption{Mean HR for simulated spectra generated using a range of temperatures $kT$ (left) or photon index $\Gamma$ (right), assuming thermal or non-thermal models, respectively, with varying $N_{\rm H}$.  Details of the simulations are given in Section \ref{sec:sim_models}.}
    \label{fig:simhr}
\end{figure}

Tables \ref{tab:apecsims} and \ref{tab:powsims} list the results of these simulations.  Each of the 274 models yielded a distribution of 1000 HR values, for which we calculated the mean, median, and $1\sigma$ uncertainty range.  Figure \ref{fig:simhr} illustrates the resulting mean HR as a function of $N_{\rm H}$ and $kT$ (for thermal spectra) or $\Gamma$ (for non-thermal spectra).  The underlying HR distributions are displayed in Figures \ref{fig:hrhist_apec} and \ref{fig:hrhist_pow} in Appendix \ref{sec:AE}.

\setlength{\tabcolsep}{8pt}
\begin{table}[]
    \centering
    \begin{tabular}{l| r| r| r| r| r| r| r}
    \hline\hline
    $N_{\rm H}$ [$10^{22}$ cm$^{-2}$] & $1.0$ & $4.0$ & $7.5$ & $11.5$ & $16.0$ & $22$ & $30$  \\
    \hline
    $kT$ [keV] & \multicolumn{7}{c}{Median HR}\\
    \hline
1.0 & -0.93$^{+0.05}_{-0.04}$ & -0.88$^{+0.07}_{-0.06}$ & -0.77$^{+0.09}_{-0.08}$ & -0.63$^{+0.11}_{-0.10}$ & -0.47$^{+0.12}_{-0.11}$ & -0.25$^{+0.12}_{-0.13}$ & 0.02$^{+0.13}_{-0.14}$ \\
2.0 & -0.67$^{+0.11}_{-0.09}$ & -0.49$^{+0.11}_{-0.13}$ & -0.28$^{+0.13}_{-0.13}$ & -0.08$^{+0.14}_{-0.13}$ & 0.12$^{+0.12}_{-0.14}$ & 0.33$^{+0.13}_{-0.12}$ & 0.52$^{+0.13}_{-0.11}$ \\
4.0 & -0.38$^{+0.13}_{-0.13}$ & -0.18$^{+0.16}_{-0.13}$ & 0.05$^{+0.13}_{-0.14}$ & 0.23$^{+0.14}_{-0.13}$ & 0.40$^{+0.13}_{-0.12}$ & 0.56$^{+0.11}_{-0.12}$ & 0.70$^{+0.10}_{-0.11}$ \\
6.0 &  -0.28$^{+0.14}_{-0.12}$ & -0.06$^{+0.13}_{-0.13}$ & 0.14$^{+0.16}_{-0.14}$ & 0.33$^{+0.14}_{-0.13}$ & 0.47$^{+0.13}_{-0.12}$ & 0.61$^{+0.12}_{-0.11}$ & 0.73$^{+0.09}_{-0.10}$ \\
11.0 &  -0.21$^{+0.13}_{-0.14}$ & 0.00$^{+0.15}_{-0.11}$ & 0.20$^{+0.15}_{-0.13}$ & 0.38$^{+0.13}_{-0.14}$ & 0.51$^{+0.12}_{-0.12}$ & 0.64$^{+0.11}_{-0.11}$ & 0.75$^{+0.10}_{-0.09}$ \\
16.0 &  -0.19$^{+0.14}_{-0.13}$ & 0.02$^{+0.16}_{-0.13}$ & 0.23$^{+0.14}_{-0.14}$ & 0.38$^{+0.14}_{-0.14}$ & 0.51$^{+0.13}_{-0.12}$ & 0.65$^{+0.11}_{-0.12}$ & 0.76$^{+0.09}_{-0.10}$ \\
21.0 & -0.17$^{+0.15}_{-0.14}$ & 0.04$^{+0.14}_{-0.14}$ & 0.23$^{+0.14}_{-0.13}$ & 0.39$^{+0.14}_{-0.15}$ & 0.51$^{+0.14}_{-0.12}$ & 0.65$^{+0.11}_{-0.11}$ & 0.76$^{+0.09}_{-0.10}$ \\
28.0 & -0.16$^{+0.16}_{-0.13}$ & 0.05$^{+0.15}_{-0.14}$ & 0.24$^{+0.15}_{-0.14}$ & 0.40$^{+0.12}_{-0.13}$ & 0.53$^{+0.12}_{-0.13}$ & 0.66$^{+0.11}_{-0.13}$ & 0.77$^{+0.10}_{-0.10}$ \\ 
36.0 &  -0.14$^{+0.13}_{-0.14}$ & 0.06$^{+0.14}_{-0.14}$ & 0.25$^{+0.14}_{-0.14}$ & 0.40$^{+0.13}_{-0.14}$ & 0.54$^{+0.12}_{-0.12}$ & 0.66$^{+0.11}_{-0.12}$ & 0.76$^{+0.10}_{-0.10}$ \\
    \hline
    \end{tabular}
    \caption{Median HR and $\pm34.1\%$ error range yielded by the thermal plasma (\texttt{APEC}) simulations.  For the models above, we assumed $Z_{\rm Fe}=Z_{\odot}$.  See Section \ref{sec:sim_models} for more.}
    \label{tab:apecsims}
\end{table}

\begin{table}[]
    \centering
    \begin{tabular}{l| r| r| r| r| r| r| r}
    \hline\hline
    $N_{\rm H}$ [$10^{22}$ cm$^{-2}$] & $1.0$ & $4.0$ & $7.5$ & $11.5$ & $16.0$ & $22$ & $30$ \\
    \hline
    $\Gamma$ & \multicolumn{7}{c}{Median HR}  \\
    \hline
    %nH= 1.0    Gamma= 
    0.0 & 0.35$^{+0.11}_{-0.11}$
&
%nH= 4.0    Gamma= 0.0 
0.49$^{+0.10}_{-0.10}$
&
%nH= 7.5    Gamma= 0.0 
0.61$^{+0.09}_{-0.10}$
&
%nH= 11.5    Gamma= 0.0 
0.69$^{+0.09}_{-0.09}$
&
%nH= 16.0    Gamma= 0.0 
0.76$^{+0.08}_{-0.08}$
&
%nH= 22.0    Gamma= 0.0 
0.83$^{+0.07}_{-0.07}$
&
%nH= 30.0    Gamma= 0.0 
0.88$^{+0.05}_{-0.07}$ \\
%nH= 1.0    Gamma= 
0.5 & 0.17$^{+0.11}_{-0.11}$
&
%nH= 4.0    Gamma= 0.5 
0.34$^{+0.11}_{-0.10}$
&
%nH= 7.5    Gamma= 0.5 
0.49$^{+0.10}_{-0.12}$
&
%nH= 11.5    Gamma= 0.5 
0.60$^{+0.11}_{-0.08}$
&
%nH= 16.0    Gamma= 0.5 
0.70$^{+0.08}_{-0.09}$
&
%nH= 22.0    Gamma= 0.5 
0.77$^{+0.08}_{-0.08}$
&
%nH= 30.0    Gamma= 0.5 
0.85$^{+0.06}_{-0.07}$ \\
%nH= 1.0    Gamma= 
1.0 & 
-0.01$^{+0.11}_{-0.11}$
&
%nH= 4.0    Gamma= 1.0 
0.19$^{+0.11}_{-0.11}$
&
%nH= 7.5    Gamma= 1.0 
0.36$^{+0.10}_{-0.11}$
&
%nH= 11.5    Gamma= 1.0 
0.49$^{+0.10}_{-0.10}$
&
%nH= 16.0    Gamma= 1.0 
0.61$^{+0.08}_{-0.10}$
&
%nH= 22.0    Gamma= 1.0 
0.71$^{+0.08}_{-0.08}$
&
%nH= 30.0    Gamma= 1.0 
0.80$^{+0.07}_{-0.08}$ \\
%nH= 1.0    Gamma= 
1.5 &
-0.20$^{+0.10}_{-0.10}$
&
%nH= 4.0    Gamma= 1.5 
0.02$^{+0.11}_{-0.11}$
&
%nH= 7.5    Gamma= 1.5 
0.21$^{+0.11}_{-0.11}$
&
%nH= 11.5    Gamma= 1.5 
0.38$^{+0.10}_{-0.10}$
&
%nH= 16.0    Gamma= 1.5 
0.51$^{+0.09}_{-0.09}$
&
%nH= 22.0    Gamma= 1.5 
0.64$^{+0.08}_{-0.09}$
&
%nH= 30.0    Gamma= 1.5 
0.76$^{+0.07}_{-0.08}$ \\
%nH= 1.0    Gamma= 
2.0 & -0.37$^{+0.10}_{-0.08}$
&
%nH= 4.0    Gamma= 2.0 
-0.15$^{+0.10}_{-0.10}$
&
%nH= 7.5    Gamma= 2.0 
0.05$^{+0.10}_{-0.11}$
&
%nH= 11.5    Gamma= 2.0 
0.24$^{+0.10}_{-0.10}$
&
%nH= 16.0    Gamma= 2.0 
0.40$^{+0.10}_{-0.09}$
&
%nH= 22.0    Gamma= 2.0 
0.55$^{+0.09}_{-0.10}$
&
%nH= 30.0    Gamma= 2.0 
0.69$^{+0.09}_{-0.08}$ \\
%nH= 1.0    Gamma= 
2.5 &
-0.51$^{+0.09}_{-0.09}$
&
%nH= 4.0    Gamma= 2.5 
-0.31$^{+0.09}_{-0.09}$
&
%nH= 7.5    Gamma= 2.5 
-0.09$^{+0.10}_{-0.10}$
&
%nH= 11.5    Gamma= 2.5 
0.11$^{+0.10}_{-0.11}$
&
%nH= 16.0    Gamma= 2.5 
0.28$^{+0.10}_{-0.10}$
&
%nH= 22.0    Gamma= 2.5 
0.46$^{+0.09}_{-0.10}$
&
%nH= 30.0    Gamma= 2.5 
0.62$^{+0.08}_{-0.09}$ \\
%nH= 1.0    Gamma= 
3.0 & 
-0.65$^{+0.08}_{-0.07}$
&
%nH= 4.0    Gamma= 3.0 
-0.46$^{+0.08}_{-0.09}$
&
%nH= 7.5    Gamma= 3.0 
-0.24$^{+0.11}_{-0.10}$
&
%nH= 11.5    Gamma= 3.0 
-0.03$^{+0.10}_{-0.10}$
&
%nH= 16.0    Gamma= 3.0 
0.16$^{+0.11}_{-0.09}$
&
%nH= 22.0    Gamma= 3.0 
0.36$^{+0.09}_{-0.10}$
&
%nH= 30.0    Gamma= 3.0 
0.54$^{+0.09}_{-0.09}$ \\
%    & & & & & & & \\
\hline
%&  \\
    \end{tabular}
    \caption{Median HR and $\pm34.1\%$ error range yielded by the non-thermal  (\texttt{POWERLAW}) simulations.  See Section \ref{sec:sim_models} for more.}\label{tab:powsims}
\end{table}

\subsection{Physical characteristics of sources corresponding to observed HR}

As Figure \ref{fig:simhr} illustrates, 
there is significant overlap between the HR distributions produced by the thermal (left) and non-thermal (right) 
models.  
In the following paragraphs, we explore what can be learned from the results of our HR simulations, as well as their limitations.

For non-thermal sources (Figure \ref{fig:simhr}; right), HR is strongly correlated with $N_{\rm H}$ in most of the parameter space explored by our simulations.  
Consequently, HR can effectively be used to constrain the spectral hardness -- \emph{where $N_{\rm H}$ is known}.  

By contrast, the thermal models (Figure \ref{fig:simhr}; left) yield HR values that appear more loosely correlated with $kT$ and $N_{\rm H}$.  
Sources with low $kT$ ($\simlt3$ keV) yield low HR values for a range of $N_{\rm H}$ spanning an order of magnitude ($10^{22}-10^{23}$ cm$^{-2}$).  Conversely, sources with high $N_{\rm H}$ ($\simgt5\times10^{22}$ cm$^{-2}$) yield HR values that are primarily dependent on $N_{\rm H}$ and almost uniform across plasma temperatures, except where the spectra are soft ($kT\simlt4$ keV).  HR appears to become saturated at $kT\sim5-10$ keV for $N_{\rm H}\simgt5\times10^{22}$ cm$^{-2}$.  Hence for thermal sources with $kT\simgt8$ keV, it becomes increasingly difficult to constrain the underlying temperature using HR, \emph{even when $N_{\rm H}$ is known}.  This is consistent with results from \cite{Nobukawa2016} (figure 8), who likewise found that the ratio of hard to broadband luminosity ($L_{X(5-10\, \rm keV)}/L_{X(2-10\, \rm keV)}$) is largely saturated at $kT\simgt5$ keV.

\begin{figure}
    \centering
    \includegraphics[width=\linewidth]{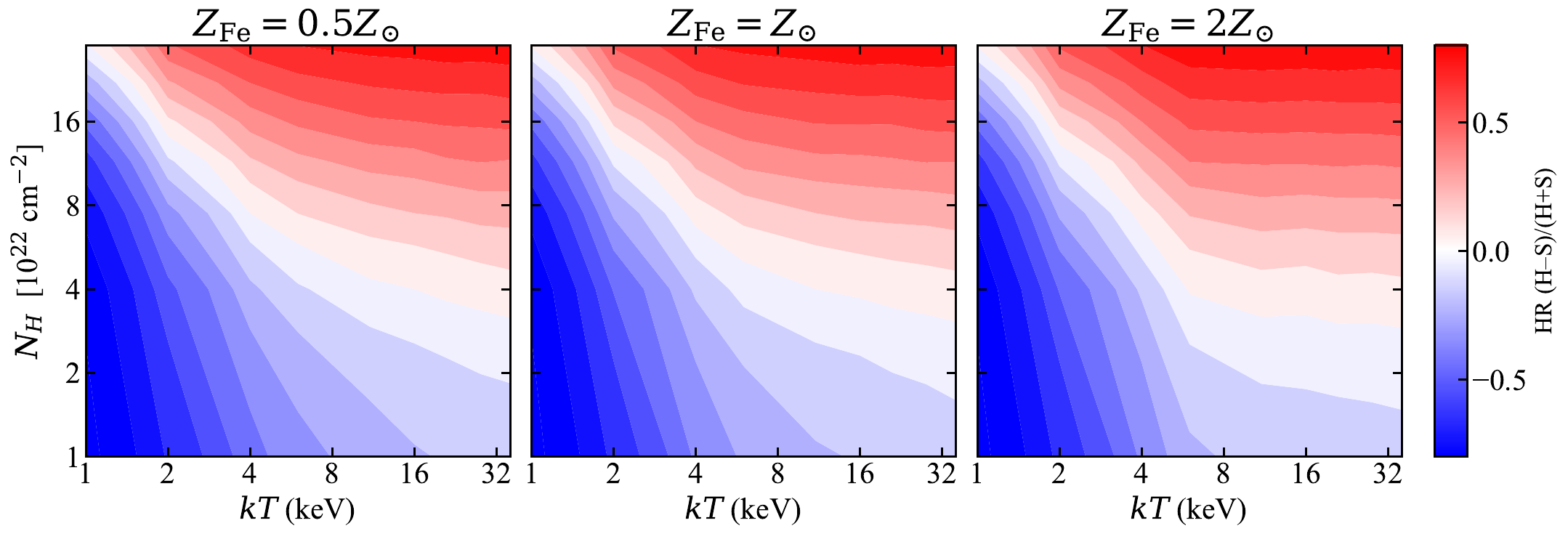}
    \caption{Average HR for thermal (\texttt{APEC}) models with a range of $N_{\rm H}$ and $kT$, assuming $Z_{\rm Fe}=0.5Z_{\odot}$ (left), $Z_{\rm Fe}=Z_{\odot}$ (center), and $Z_{\rm Fe}=2Z_{\odot}$ (right).  Varying $Z_{\rm Fe}$ within the range of $(0.5-2)\times Z_{\odot}$ does not significantly change the corresponding HR.}
    \label{fig:3fe}
\end{figure}

Perhaps more surprisingly, our results indicate that -- at least for thermal plasma models -- the \emph{average} HR shows very little dependence on the Fe abundance for a range of $Z_{\rm Fe}$ between $0.5-2\, Z_{\odot}$, despite the Fe emission falling squarely in the hard band (Figure \ref{fig:3fe}).  This indicates that the hard source concentration we observe in the NSD region is not simply a result of higher $Z_{\rm Fe}$ for the sources within that region.  Note, however, that we do not disprove the presence of sources with higher $Z_{\rm Fe}$; we simply show that HR is not sensitive enough to $Z_{\rm Fe}$, at least within the range of values commonly assumed for sources in the NSD.  Other methods can (and do) indicate that a population with $Z_{\rm Fe}>Z_{\odot}$ may in fact be present in that region (cf. \cite{Anastasopoulou2023} and references therein).

While we cannot positively identify most individual sources based on HR alone, the HR \emph{distributions} of the thousands of point sources in our sample are nevertheless illuminating, particularly when we utilize the observed line-of-sight (LoS) column density $N_{\rm H}$ to mitigate the degeneracy between X-ray absorption and intrinsic spectral hardness.  (This assumes, of course, that the sources in question are not located in the foreground, with lower corresponding source-specific $N_{\rm H}$.)  Other studies indicate that the fraction of foreground sources in our sample is relatively small (e.g., \cite{Bao2025} found eROSITA All-Sky Survey (eRASS1) counterparts -- which would be expected for most X-ray sources from our sample if they were located in the solar neighborhood -- for $<1\%$ of the sources in this survey).  We therefore proceed under the assumption that for most sources in our sample, the observed LoS $N_{\rm H}$ is a reliable approximation for the column density towards the source.   

Hereafter, it becomes important to consider $N_{\rm H}$ when using our simulation results to connect the observed HR distributions (Section \ref{sec:hr}) to physical source properties:

Outside the NSD ($|l|\simgt1.5^{\circ}-2^{\circ}$), LoS $N_{\rm H}$ ranges primarily between $(2-4.5)\times10^{22}$ cm$^{-2}$.  Within that range, 
\emph{thermal} sources with HR $\simlt-0.2$ ($\sim32\%$ of the sample) correlate with temperatures $kT<5$ keV (e.g., active binaries and non-magnetic CVs), while those with HR $>-0.2$ have $kT>5$ keV (e.g. magnetic CVs).  \emph{Non-thermal} sources have HR strongly and $\sim$linearly correlated with spectral index $\Gamma$.  Most sources outside the NSD with HR $>0$ ($\sim55\%$ of the sample)
likely either fall in the non-thermal category (e.g., MSPs), or have additional source-specific absorption (e.g. IPs with strong partial covering by accretion curtains), as HR from purely thermal spectra in our $2-12$ keV band saturates for higher $kT$ and likely would not exceed HR $=0$.

Within the NSD ($|l|\simlt1.5^{\circ}$), LoS $N_{\rm H}$ varies considerably, with some regions exhibiting $N_{\rm H}$ between $(4-8)\times10^{22}$ cm$^{-2}$ and others (particularly in the CMZ) ranging between $(6-12)\times10^{22}$ cm$^{-2}$ (Figure \ref{fig:hrvnh_den}).  In small regions along dense molecular clouds (e.g., Sgr B2), $N_{\rm H}$ is even higher, exceeding $1.2\times10^{23}$ cm$^{-2}$.  \emph{Thermal} sources with $kT$ between $1-36$ keV and $N_{\rm H}<10^{23}$ cm$^{-2}$ yield HR values up to $\sim0.4$. Only sources with $N_{\rm H}>10^{23}$ cm$^{-2}$ \emph{and} $kT\simgt2$ keV should result in higher HR values, unless additional source-intrinsic absorption is present. 
\emph{Non-thermal} sources exhibit a wider range of HR, spanning $-0.6\simlt\rm HR \simlt 0.6$.  HR is strongly correlated with both $\Gamma$ and $N_{\rm H}$.  At very high $N_{\rm H}$ ($>10^{23}$ cm$^{-2}$), HR can exceed 0.6 for very hard sources ($\Gamma\simlt1-1.5$, depending on $N_{\rm H}$) and becomes less sensitive to spectral hardness.  HR values $>0.8$ require both very high absorption (e.g., $N_{\rm H}>2\times10^{23}$ cm$^{-2}$) and very hard spectra (e.g. $\Gamma<1$).  $\sim15\%$ of our sample exhibits such high HR.  Of those sources, nearly all ($\sim90\%$) are located in regions with LoS $N_{\rm H}<2\times10^{23}$ cm$^{-2}$, indicating additional source-intrinsic absorption.

A more detailed examination of the possible source types and their spatial distributions is presented in Section \ref{sec:discuss}.  \\

\section{Distributions of hard and soft sources inside and outside the NSD}\label{sec:distr}

For the following analysis, we consider the PN and MOS datasets in combination. For sources detected by both the PN and MOS instruments, we adopt the HR values derived from the PN data.

\subsection{Flux distribution and completeness} \label{sec:comp}

\begin{figure}
    \centering
    \includegraphics[width=0.6\linewidth]{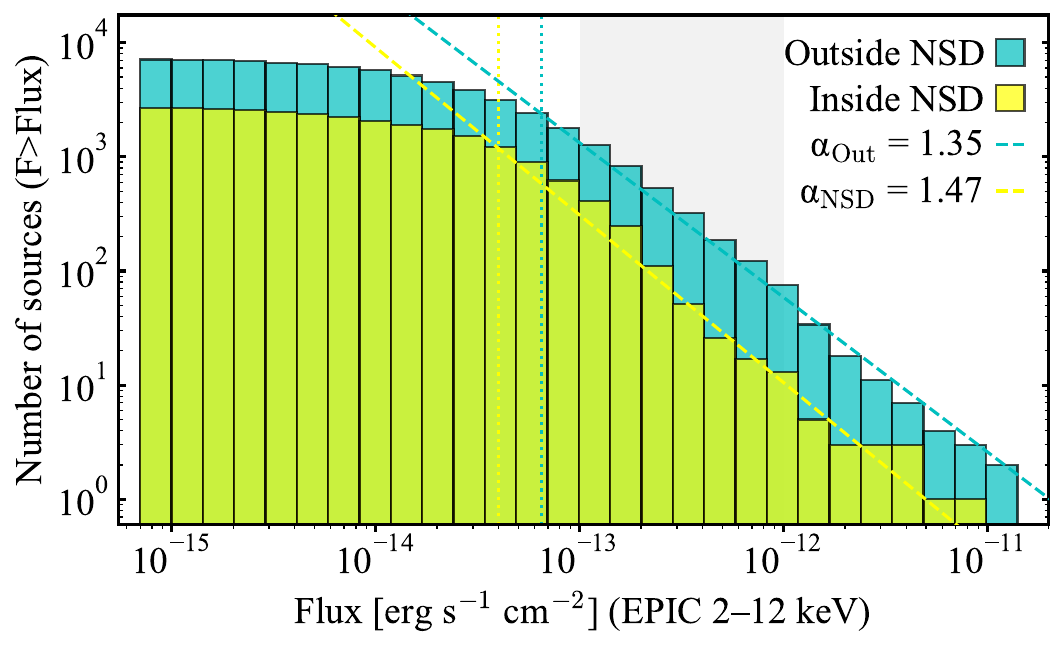}
    \caption{log$N$-log$S$ distributions for sources inside (yellow) and outside (cyan) the NSD region.  Dotted lines mark the completeness limits of the distributions.  
    Both distributions fit to a power-law (dashed lines), with $\alpha \sim1.35-1.5$. 
    The shaded region indicates the range where the lines were fit.}
    \label{fig:fxdist}
\end{figure}

We estimate the log$N$-log$S$ distributions for the X-ray sources inside and outside the NSD (Figure \ref{fig:fxdist}).  Both distributions appear to have a slope of $\alpha\sim1.35-1.5$ (dashed lines), consistent with the results from \cite{Muno2009}.  Both flux distributions appear to be complete to $f_X\approx10^{-13}$ \fluxcgs\ (dotted lines), and at least $80\%$ complete at $f_X\sim5\times10^{-14}$ \fluxcgs.

\begin{figure}
    \centering
    \includegraphics[width=\linewidth]{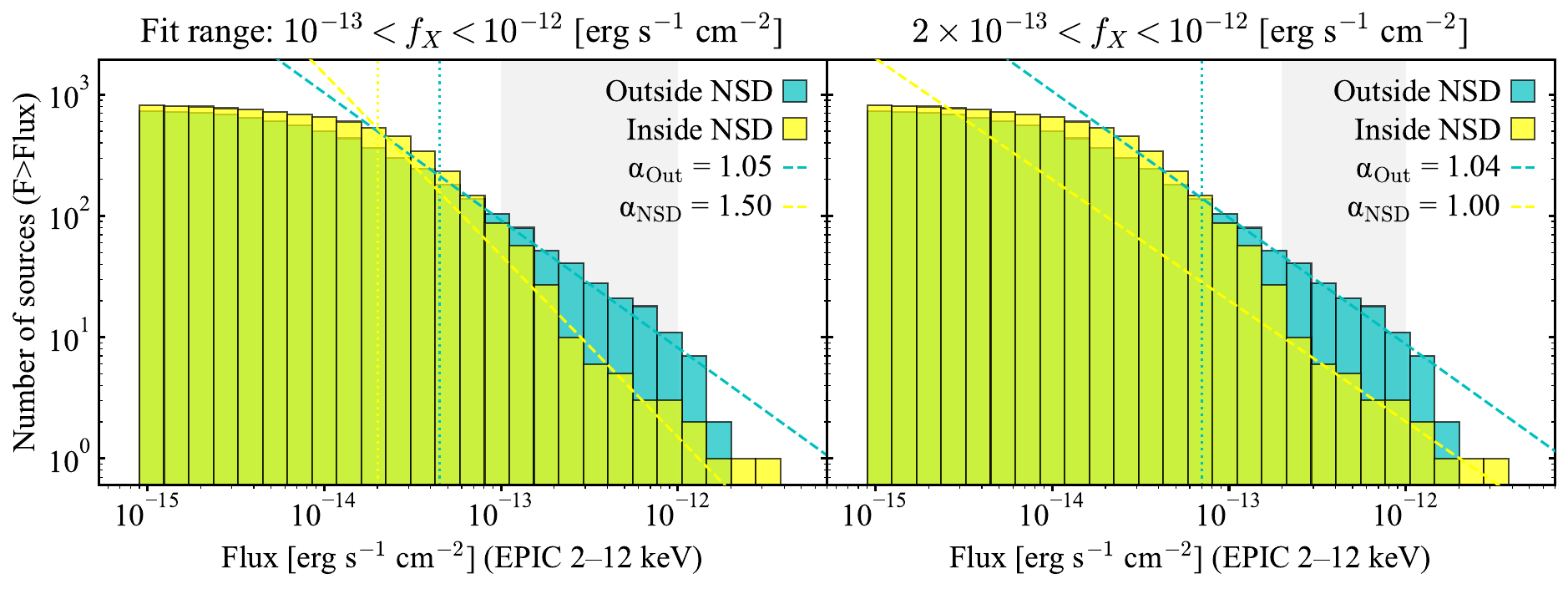}
    \caption{log$N$-log$S$ distributions for sources from a uniform exposure range inside (yellow) and outside (cyan) the NSD region, fit over a narrow flux range ($2\times10^{-13}<f_X<10^{-12}$ \fluxcgs; right) and a wider flux range ($10^{-13}<f_X<10^{-12}$ \fluxcgs, left).  Dashed lines mark the best-fit line; shaded regions indicate the range over which the lines were fit; dotted lines mark the flux where the distributions appear complete.  The best-fit slope $\alpha$ is dependent on the fitting range.  See Section \ref{sec:comp} for more. }
    \label{fig:fxdist_expfilt}
\end{figure}

As before (Section \ref{subsec:test_hr}), we consider the effects of unequal exposures on our analysis.  Regions with longer exposures will be more sensitive to faint sources, while those with short exposures will favor bright sources, potentially biasing the distribution.  We therefore once again select only sources with 
exposures ranging between $22-53$ ks for PN and $60-138$ ks for MOS, where the respective exposure distributions inside and outside the NSD are equivalent (Figure \ref{fig:ctrt_dist}).  
When we evaluate the flux distributions after applying the exposure filters, we find that the log$N$-log$S$ appears steeper inside the NSD and no longer $\sim$linear below $\sim2\times10^{-13}$ \fluxcgs\ (Figure \ref{fig:fxdist_expfilt}).  This is likely a consequence of the model parameters used to extrapolate the DR14 catalog flux ($N_{\rm H}=3\times10^{20}$ cm$^{-2}$, $\Gamma=1.7$); with much higher $N_{\rm H}$ and harder average $\Gamma$ in our dataset, this model would yield systematically fainter fluxes in the inner region\footnote{Although $N_{\rm H}$ in our survey region outside the NSD is also higher than the modeled $N_{\rm H}$ applied to DR14, the difference is not as stark compared to the NSD, and as shown in Section \ref{sec:hr}, the sources are also softer on average and more consistent with the modeled photon index $\Gamma=1.7$ (Section \ref{sec:sims}).  The effect on the NSD flux distribution is also less pronounced without the exposure filter, possibly because hundreds more faint sources from observations with exposures $>100$ ks (Figure \ref{fig:ctrt_all}) are included, smoothing out the distribution. }, shifting the distribution towards lower values.  We also find that the best-fit slope (and the derived completeness limit) is somewhat dependent on the choice of the range of fluxes over which it is fit: using a wider range to fit the line yields a shallower slope $\alpha$ for the NSD distribution with its fainter tilt (Figure \ref{fig:fxdist_expfilt}, left).  Given these biases,  our measured values for $\alpha$ are assumed to be approximate (see Appendix \ref{sec:AC} for more details on how selection parameters affect $\alpha$, including a breakdown of the flux distributions for hard vs soft sources).  The range of $\alpha=1.0-1.5$ is, however, consistent with the log$N$-log$S$ range measured by \cite{Muno2009} and \cite{Hong2009} in several fields around
the Galactic Center, some of which overlap our survey.  
A future spectroscopic study of the point sources in our sample could yield more accurate fluxes and log$N$-log$S$ distributions.

For our analysis in the following sections, we apply a minimum flux threshold of either $f_X=10^{-13}$ \fluxcgs\ (for a smaller but less biased sample), or $f_X=5\times10^{-14}$ \fluxcgs\ (for a potentially \emph{slightly} biased, larger sample with smaller statistical errors), and sometimes both.

\subsection{Spatial distributions and scale heights of hard and soft sources} \label{sec:shdist}

Having determined that there are at least two distinct source populations that dominate within different regions of the central Galactic disk, we investigated the spatial distributions of these two groups over the full survey region.  We begin by dividing the sources into "hard" and "soft" groups, with a dividing threshold of HR$=0.4$ for PN and HR$=0.25$ for MOS.  These thresholds were determined by the intersection of the respective bimodal HR distributions, as shown in Figure \ref{fig:hr3}.  We note that many sources with HR below those thresholds might traditionally be considered quite hard; in this study, we single out only sources with atypically high HR as "hard" and less hard sources as "soft", even where the latter would not ordinarily be classified as soft X-ray sources.  To minimize biases introduced by varying exposure, background, and source confusion over the survey area, we include only sources with flux $f_X>5\times10^{-14}$ \fluxcgs\ in the following analysis.  

\begin{figure}
\centering
\includegraphics[width=1.\textwidth]{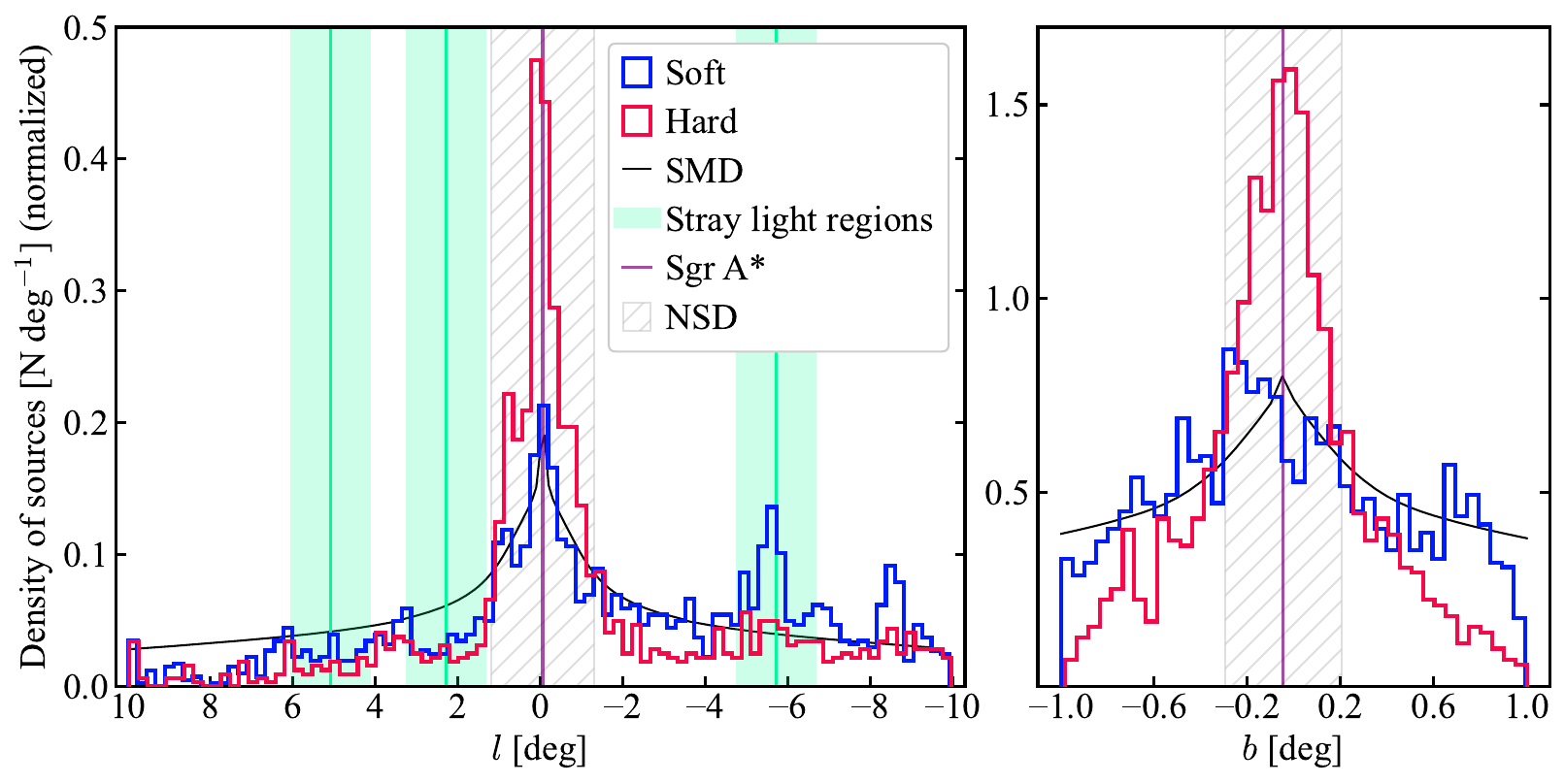}
\caption{Distributions of hard (red) and soft (blue) X-ray sources as a function of Galactic longitude $l$ (left) and latitude $b$ (right).  Green columns demarcate the regions with stray-light contamination, resulting in a deficit of sources (as shown in Figure \ref{fig:exp}).  The threshold dividing hard and soft sources is HR$=0.4$ for PN and HR$=0.25$ for MOS.  The location of \sgra\ is marked by a purple line.  Inside the NSD (hatched area), harder sources are more sharply concentrated towards \sgra\ compared to the soft sources and the analytical SMD.   
See Section \ref{sec:shdist} for more.  }
\label{fig:sh_dist1}
\end{figure}

We first observe the distributions of the hard and soft sources over the range of Galactic longitudes $|l|\simlt10^{\circ}$ by plotting their respective density distributions in a histogram normalized to the total number of sources in each group.  As Figure \ref{fig:sh_dist1} (left) shows, both distributions peak within $|l|\simlt1^{\circ}$.  However, the hard sources (red) have a distinctly sharper peak compared to their soft counterparts (blue), with a normalized peak "height" nearly double that of the soft distribution.  Outside of $|l|\simlt1^{\circ}$, the hard and soft sources appear similarly distributed.  There is a noticeable dearth of sources at longitudes corresponding to regions subjected to stray light contamination (highlighted in green), which have a radius of $\sim1.2^{\circ}$.  Incomplete survey coverage at $l>7^{\circ}$ -- rather than an intrinsic lopsided distribution -- is most likely responsible for the lower number density of sources at eastern longitudes.

We then evaluated the distributions of hard and soft sources over Galactic latitudes $|b|<1^{\circ}$.  Here the difference between the two distributions is similarly pronounced, with the hard sources sharply peaked at $-0.2\simlt b\simlt0.1$.

A K--S test indicates that the distributions of hard and soft sources over Galactic longitude/latitude are distinct at a level $>12\sigma/9\sigma$, with $p$-values of $2\times10^{-35}$/$3\times10^{-20}$, respectively.

\begin{figure}
\centering
\includegraphics[trim=0.25cm 0.5cm 0.25cm 0.25cm, width=1.\textwidth]{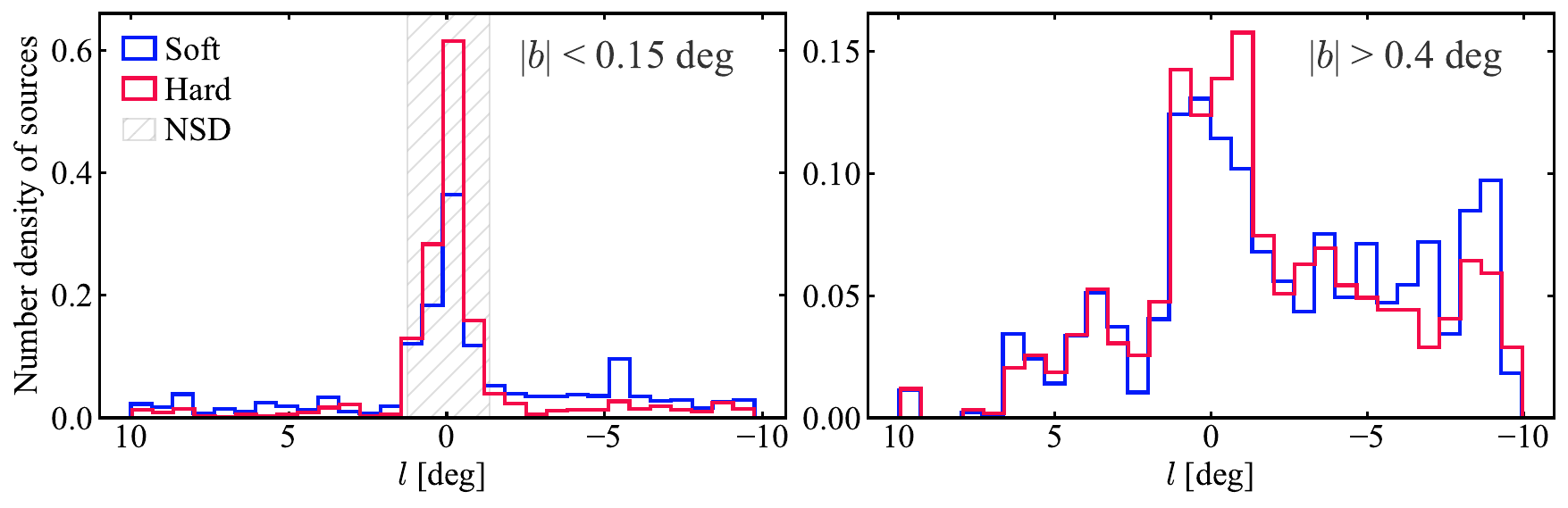}
\caption{Distributions of hard (red) and soft (blue) X-ray sources as a function of Galactic longitude $l$, within Galactic latitudes corresponding to the NSD ($|b|<0.15^{\circ}$, left) and outside the NSD ($|b|>0.4^{\circ}$, right).  The threshold dividing hard and soft sources is HR$=0.4$ for PN and HR$=0.25$ for MOS.  While the distribution of hard sources is much more sharply peaked -- compared to the soft distribution -- within the NSD, outside of the central $0.4^{\circ}$, the hard and soft distributions are similar.  
See Section \ref{sec:shdist} for more.}
\label{fig:sh_dist3}
\vspace{0.75cm}
\includegraphics[trim=0.25cm 0.5cm 0.25cm 0.25cm, width=1.\textwidth]{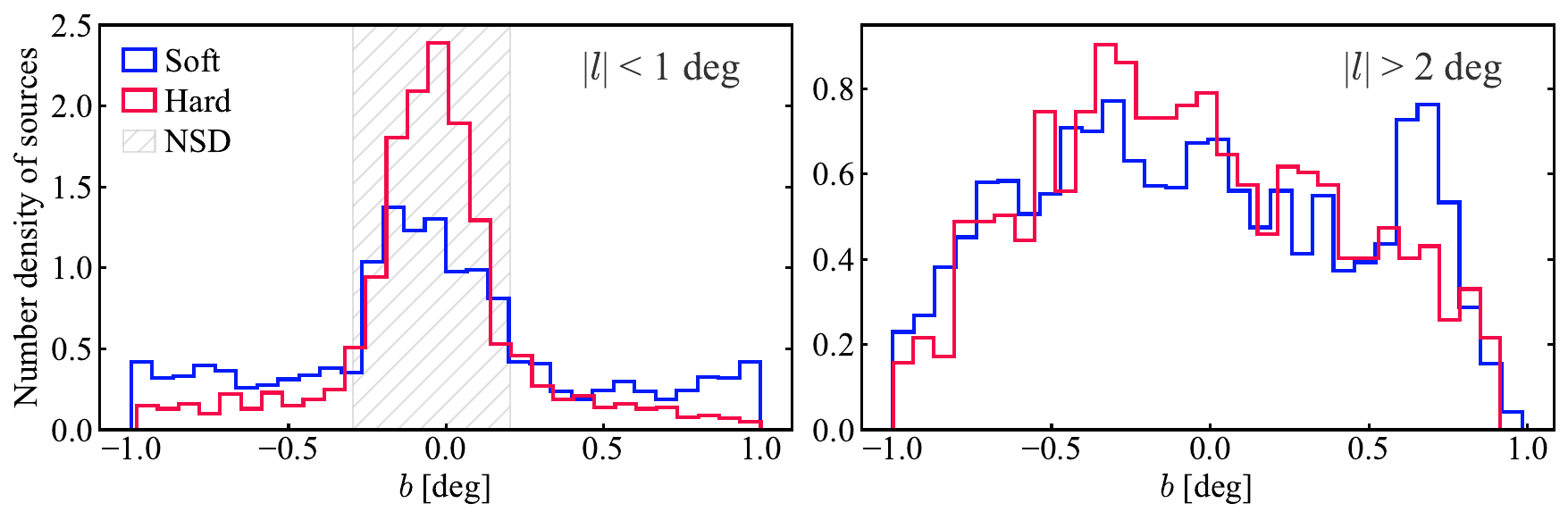}
\caption{Distributions of hard (red) and soft (blue) X-ray sources as a function of Galactic latitude $b$, within the region including the NSD ($|l|<1^{\circ}$, left) and outside the NSD ($|l|>2^{\circ}$, right).  The threshold dividing hard and soft sources is HR$=0.4$ for PN and HR$=0.25$ for MOS.  As is the case for the longitudinal distributions shown in Figure \ref{fig:sh_dist3}, the hard sources are more centrally concentrated within the NSD compared to the soft sources, while at $|l|>2^{\circ}$, the hard and soft distributions are similar.  
See Section \ref{sec:shdist} for more.}
\label{fig:sh_dist4}
\end{figure}

We then test our results by evaluating how the hard and soft distributions differ within and outside the NSD regions ($|l|\simlt1.3^\circ$, $|b|\sim0.3^\circ$), using conservative limits that exclude the surrounding "borders".  For this analysis, we first select a region along the center of the Galactic plane ($|b|<0.15^{\circ}$) and at higher latitudes ($|b|>0.4^{\circ}$).  We compare the  longitudinal distributions of hard and soft sources within these separate regions, finding that the distributions along the center are much more divergent -- with the hard sources sharply peaked -- compared to that at higher latitudes, where the hard and soft distributions appear similar (Figure \ref{fig:sh_dist3}).  Likewise, we split 
the sources at the central Galactic longitude $|l|<1^{\circ}$ and outer regions $|l|>2^{\circ}$.  Once again we find that in the central $2^{\circ}$ longitude, the distributions along Galactic latitude sharply diverge -- as before, the hard source distribution is much more concentrated towards $|b|\simlt0.2$, while in the outer regions, the hard and soft distributions are largely consistent (Figure \ref{fig:sh_dist4}).

\smallskip

Finally, we compare the distributions of hard and soft sources to an analytical model of the stellar mass distribution (SMD) at the Galactic Center, comprising four components: the Galactic disk, the Galactic bar, the NSD, and the nuclear star cluster (NSC).  The model, derived from \cite{Chatzopoulos2015, Sormani2022a, Sormani2022b}, is described in \cite{Anastasopoulou2023} (Model 2).  For this analysis, we binned the SMD along the central $\sim20^{\circ}\times2^{\circ}$, matching the \xmm\ survey region.  For the Galactic longitude profile, we utilized bins of width and height $(\delta l, \delta b) = (0.1^{\circ}, 2^{\circ})$; for Galactic latitude, we applied $(\delta l, \delta b) = (19^{\circ}, 0.05^{\circ})$.  Figure \ref{fig:sh_dist1} shows the binned SMD data alongside the normalized Galactic longitude (left) and latitude (right) distributions of the hard (HR $>0.4$; red) and soft (HR $<0.4$; blue) sources.  The SMD model was normalized to fit the hard and soft distributions outside the NSD (i.e., at $|l|>1.5^{\circ}$).  It appears that the SMD traces the soft source distribution much better than it does the hard sources, indicating that the soft X-ray sources may be related to the overall stellar population.  By contrast, the hard sources are far more concentrated towards \sgra.  
An enhanced view of the longitudinal distributions within $|l|\simlt5^{\circ}$, as well as the residuals after subtracting the SMD from the data, can be found in Figure \ref{fig:smd_zoom}.

\begin{figure}
\centering
\includegraphics[width=0.9\textwidth]{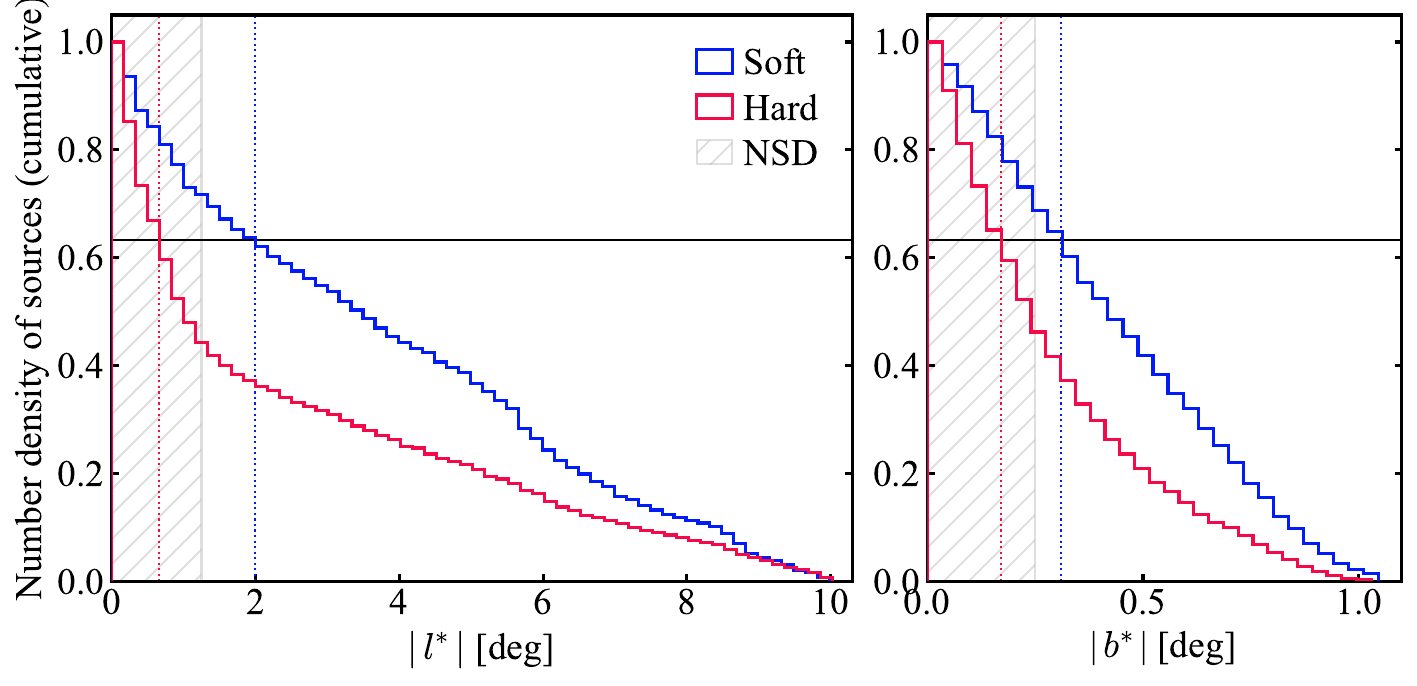}
\caption{Cumulative density profiles of hard (red) and soft (blue) X-ray sources as a function of \sgra-centered Galactic longitude $l^*$ (left) and latitude $b^*$ (right).  The threshold dividing hard and soft sources is HR$=0.4$ for PN and HR$=0.25$ for MOS.  Horizontal black lines denote the location where the density is reduced from the peak by a factor of $1/e$; we take the locations where the lines intersect the density distributions as the scale heights (vertical dotted lines).  
See Section \ref{sec:shdist} for more.  }
\label{fig:sh_dist2}
\end{figure}

For a more quantitative approach to the distributions of the respective hard and soft source populations, we evaluate their "scale height" $h$ -- i.e. the distance from the center at which the source density drops by a factor of $1/e$ from its peak value.  For the purposes of this analysis, we define \sgra-centered coordinates $l^*, b^*=l+l_{\rm Sgr A^*}, b+b_{\rm Sgr A^*}$, such that the distributions of the sources in %longitude and latitude
$l^*$ and $b^*$ are defined with respect to \sgra.  As Figure \ref{fig:sh_dist2} (left) shows, the scale height in longitude of the soft sources, ($h_{l,\rm s}=2^{\circ}$ [$\sim280$ pc]) is more than double that of the hard sources ($h_{l,\rm h}=0.66^{\circ}$ [$\sim90$ pc]).  The hard distribution is much more steep at $|l|\simlt1.3^{\circ}$, leveling off outside the NSD (hatched region).  Similarly, the scale height in latitude is much higher for the soft distribution ($h_{b,\rm s}=0.31^{\circ}$ [$\sim44$ pc]) vs the hard sources ($h_{b,\rm h}=0.17^{\circ}$ [$\sim24$ pc]); the hard distribution appears to level off somewhat outside the NSD ($|b|\sim0.3^\circ$), while the slope of the soft sources appears constant out to $|b|=1^{\circ}$ (Figure \ref{fig:sh_dist2}, right). 

However, these derived scale heights should be taken strictly as upper limits.  As Figure \ref{fig:badsources} demonstrates, the number of sources in the DR14 catalog that are flagged as potentially spurious or having unreliable parameters (which were removed from our analysis) is higher than the number of "clean" sources remaining in our sample in the region $-0.2^\circ<l<0.1^\circ$ (within $0.15^\circ$, or $\sim21$ pc, of \sgra), a consequence of the high background and source crowding in that region\footnote{The number of sources missing from the NSC region due to source confusion is potentially even more significant, as indicated by the dearth of sources in the central $\simlt0.1^{\circ}$ illustrated in Figure \ref{fig:badsources} (inset).}.  Even if we assume that none of the "potentially spurious" sources are real (a somewhat unlikely supposition), the sources with unreliable parameters alone would raise the number of sources within $|l^*|<0.15^\circ$ by nearly $40\%$ (whereas their contribution within the survey overall is $\sim18\%$, yielding a "net" increase of $\sim20\%$ in the center.)  The true scale heights are therefore even smaller.  While we cannot reliably determine the HR for the sources with unreliable parameters, their steep density profile is more consistent with the hard source distributions (which also appear to be most affected by confusion in the central $\sim0.1^{\circ}$, as indicated by Figure \ref{fig:smd_zoom}). 

Table \ref{tab:scales} lists the scale heights observed for the NSD (\cite{Launhardt2002, Nishiyama2013, Sormani2022b, Nezabudkin2026}) and the overlapping Galactic Center diffuse emission (\cite{Uchiyama2013, Heard2013, Yamauchi2016}) in previous studies.  Our results are broadly consistent with past estimates, which have varied substantially -- perhaps because they probed different populations. \\

\setlength{\tabcolsep}{5pt}
\begin{table}[]
    \centering
    \begin{tabular}{l| l |c | c}
    \hline\hline
    Study & Data & Scale height $l$ & Scale height $b$ \\
    \hline
       \cite{Launhardt2002} & IRAS/COBE DIRBE (NIR) & $230\pm20$ pc (radius) & $45\pm5$ pc ($\sim0.32^{\circ}$) \\
        \cite{Nishiyama2013} & IRSF/SIRIUS (NIR) & -- & $45$ pc ($0.32^{\circ}\pm0.02^{\circ}$) \\
        \cite{Uchiyama2013} & \suzaku\ (Fe XXV) & $89$ pc ($0.63^{\circ}\pm0.03^{\circ}$) & $38$ pc ($0.27^{\circ}\pm0.01^{\circ}$) \\
        \cite{Uchiyama2013} & \suzaku\ ($2.3-5$ keV) & $89$ pc ($0.63^{\circ}\pm0.03^{\circ}$) & $59$ pc ($0.42^{\circ}\pm0.04^{\circ}$) \\
        \cite{Uchiyama2013} & \suzaku\ ($5-8$ keV) & $102$ pc ($0.72^{\circ}\pm0.06^{\circ}$) & $44$ pc ($0.31^{\circ}\pm0.02^{\circ}$) \\
         \cite{Heard2013} & \xmm\ ($6-10$ keV% unresolved
         ) & -- & 44 pc ($0.31^{\circ}$) \\
         \cite{Yamauchi2016} & \suzaku\ (6.7 keV Fe XXV) & -- & $36\pm3$ pc ($0.26^{\circ}\pm0.02^{\circ}$) \\
         \cite{Yamauchi2016} & \suzaku\ (5-8 keV) & -- & $35\pm3$ pc ($0.25^{\circ}\pm0.02^{\circ}$) \\
         \cite{Sormani2022b} & VIRAC2 (NIR) & $88.6^{+9.2}_{-6.9}$ pc ($0.63^{\circ}$) & $28.4^{+5.5}_{-5.5}$ pc ($0.2^{\circ}$) \\
        \cite{Nezabudkin2026} & ART-XC & $100$ pc ($0.71^{\circ}$) & $20$ pc ($0.14^{\circ}$) \\
         This work & \xmm\ soft point sources  & 280 pc ($2^{\circ}$) & 44 pc ($0.31^{\circ}$) \\
         This work & \xmm\ hard point sources  & $<90$ pc ($0.66^{\circ}$) & $<24$ pc ($0.17^{\circ}$) \\
         \hline
    \end{tabular}
    \caption{Table listing scale height estimates for the NSD or GCXE.}
    \label{tab:scales}
\end{table}

\section{Discussion} \label{sec:discuss}

The analysis completed in Section \ref{sec:hr} indicates that the central $\sim2.5^{\circ}\times0.5^{\circ}$ of the Galaxy -- a region corresponding to the NSD -- is home to a unique population of hard X-ray sources that is distinct from the X-ray source population in other parts of the Galactic disk.  In this Section, we explore the possible causes for this observed contrast in hardness, and probe the underlying source types that could produce the observed distributions.  A list of Galactic structures referred to in this section -- and their spatial extents -- can be found in Table \ref{tab:abbr}.

As described in the Introduction of this work, the origin and composition of the GCXE has long been under debate. Several studies over the past two decades have shown an excess of Fe XXV emission -- with respect to both the surrounding GRXE (e.g., \cite{Uchiyama2011}) and the overlapping NIR stellar density profile (e.g., \cite{Nishiyama2013}) -- corresponding to the footprint of the NSD. While there is broad agreement on the higher X-ray emissivity within that region, interpretations vary widely; large-scale studies examining the diffuse Fe-K emission have often favored a diffuse origin (\cite{Koyama2007, Uchiyama2011, Uchiyama2013, Nishiyama2013}), whereas more confined studies of X-ray point sources typically concluded that a point source origin is more likely (\cite{Heard2013, Xu2019}).\footnote{There are notable exceptions where diffuse emission was similarly interpreted as originating from point sources, e.g. \cite{Anastasopoulou2023} used a study of the diffuse Fe XXV emission detected by our \xmm\ Heritage Survey to show that point sources with Fe abundances higher than those in the Galactic bulge can account for most of the Fe XXV "excess".  \cite{Hailey2016} likewise showed that the $20-40$ keV central hard X-ray emission (CHXE) -- detected by \nustar\ within the footprint of the NSC -- is consistent with a population of hard IPs.}  However, prior studies of point sources in relation to the GCXE were invariably 
limited in scale \citep{Hong2009, Muno2009, Heard2013, Zhu2018, Xu2019, Schmitt2022}.  This study presents the first comprehensive investigation of X-ray point sources spanning not only the full extent of the GCXE and NSD, but also extending out to the Galactic disk up to $|l|<10^{\circ}$.  The \xmm\ Heritage Survey allows us to evaluate the X-ray point source distributions and how they vary in different regions, analysis that was heretofore limited to diffuse X-ray emission. 

\begin{figure}
    \centering
    \includegraphics[width=\linewidth]{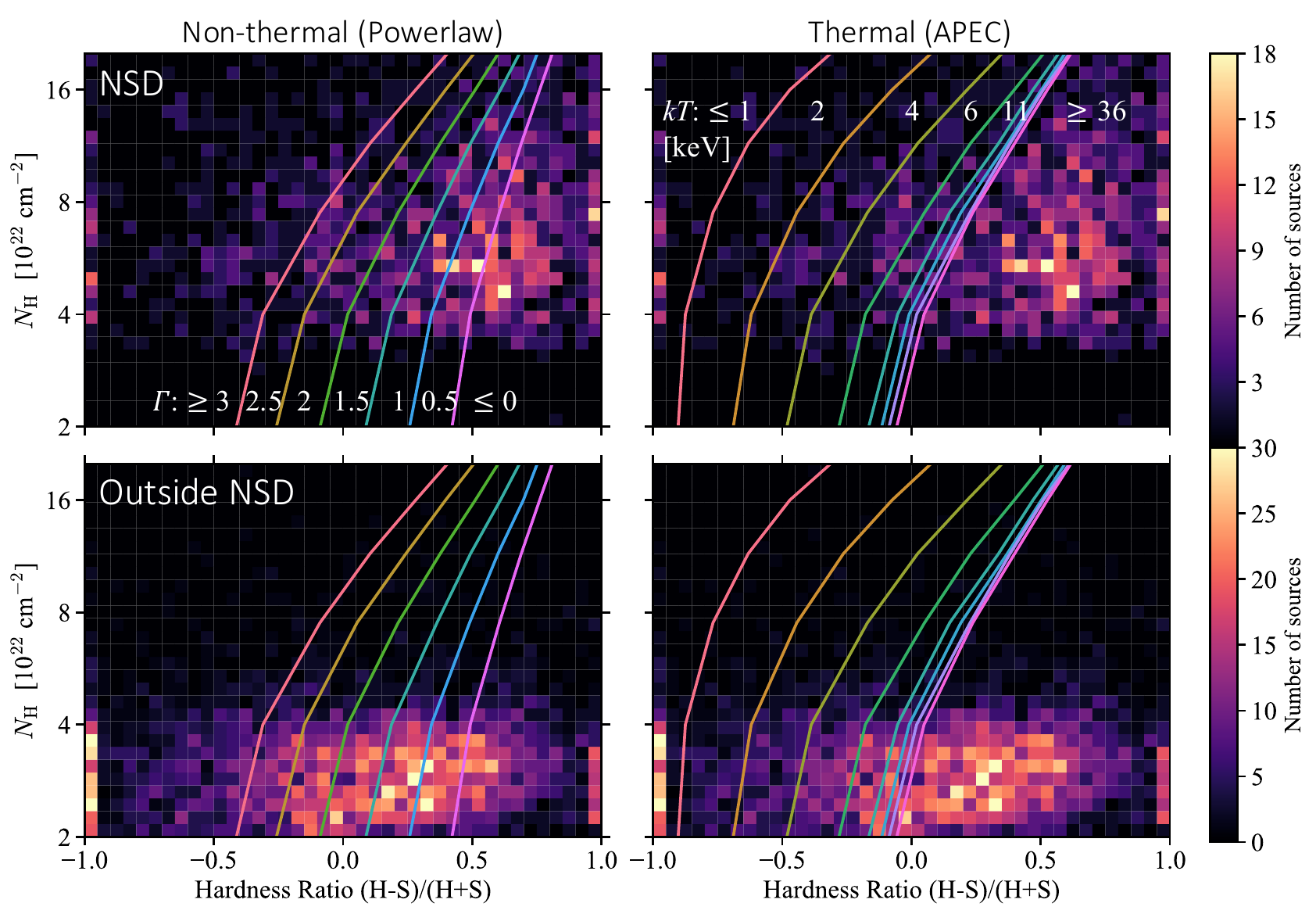}
    \caption{Density of PN sources as a function of observed HR and $N_{\rm H}$, within the NSD (top), outside the NSD (bottom), with contours of corresponding photon index $\Gamma$ (left), and contours of corresponding temperature $kT$ (right) overlaid.  Many of the sources at HR $>0$ are likely either non-thermal sources, or IPs with additional hardening due to obscuration by accretion curtains.  See Section \ref{subsec:mCVs} for more.}
    \label{fig:hrvnh_den}
\end{figure}

\begin{figure}
    \centering
    \includegraphics[width=0.85\linewidth]{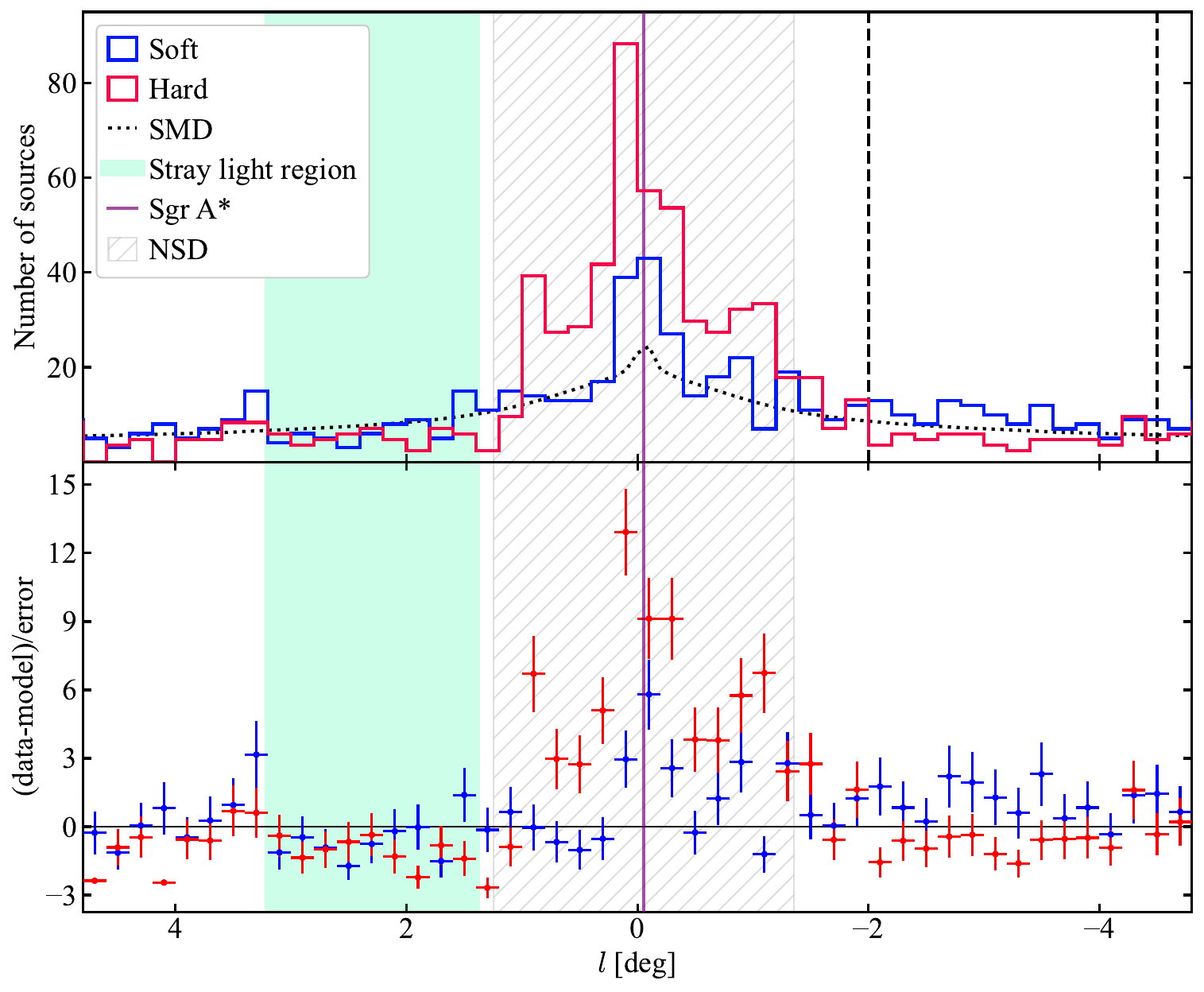}
    \caption{Top: Histogram showing the density distribution of hard (red) and soft (blue) point sources along $|l|<4.8$ with $f_X>10^{-13}$  \fluxcgs.  
    The threshold dividing hard and soft sources is HR$=0.4$ for PN and HR$=0.25$ for MOS.  Black vertical dashed lines bracket the region where the SMD model was normalized to the data.  
    Bottom: residuals ([data-model]/error)  
    highlight the excess of hard sources in the NSD (hatched area).  }
    \label{fig:smd_zoom}
\end{figure}

Below we evaluate the composition of the X-ray sources within and outside the NSD, utilizing the observed HR distributions and LoS $N_{\rm H}$ to draw inferences about their physical characteristics, based on the results of the HR simulations in Section \ref{sec:sims}.  Table \ref{tab:phys} provides a summary of the underlying source populations in different regions based on the HR simulations, which are also illustrated in Figure \ref{fig:hrvnh_den}.  The percentages provided are based on the PN sample, as the simulations were conducted using PN response files.  However, we expect comparable results for the MOS sample, since their HR distributions are similar to the PN sample after adjusting for differences in effective area (as demonstrated in Figures \ref{fig:hr3}$-$\ref{fig:degbin_b}.)   \\

\setlength{\tabcolsep}{5pt}
\renewcommand{\arraystretch}{1.2}
\begin{table}[]
\begin{center}
\caption{Physical models corresponding to observed HR.  The percentage of sources within the specified ranges of HR and $N_{\rm H}$ (based on the PN sample) are provided in parentheses.  (The percentages are expected to be similar for MOS after accounting for differences in effective area.)}\label{tab:phys}
\begin{tabular}{c | c | c}
\hline\hline
 & \textbf{Within the NSD} & \textbf{Outside the NSD} \\
 & ($|l|\simlt1.5^{\circ}$, $|b| < 0.2^{\circ}$) & ($|l|\simgt1.5^{\circ}-2^{\circ}$ or $|b| \simgt0.2^{\circ}-0.3^{\circ}$) \\
 & $N_{\rm H} \simgt(4-8)\times10^{22}$ cm$^{-2}$ & $N_{\rm H} \sim(3-5)\times10^{22}$ cm$^{-2}$ \\
\hline
\multirow{3}{2.5cm}{\textit{Non-thermal} sources (\tt{POWERLAW})} & \multirow{3}{8.2cm}{Wider range of HR ($-0.4\simlt\rm HR \simlt 0.6$), strongly correlated with $\Gamma$ and $N_{\rm H}$ ($\sim54\%$ of the sample); for $N_{\rm H}>10^{23}$ cm$^{-2}$, HR $>0.6$ for $\Gamma\simlt1-1.5$ ($\sim11\%$).} &  \multirow{3}{6.cm}{HR strongly correlated with $\Gamma$; may include many sources with HR $>0$ ($\sim55\%$ of the sample).} \\
  & & \\
  & & \\
  \hline
\multirow{3}{2.5cm}{\textit{Thermal} sources (\tt{APEC})} 
  & \multirow{3}{8.2cm}{$1<kT<36$ keV and $N_{\rm H}<10^{23}$ cm$^{-2}$ correspond to HR up to $\sim0.4$ ($\sim36\%$ of the sample); $N_{\rm H}>10^{23}$ cm$^{-2}$ \emph{and} $kT\simgt2$ keV yield HR $>0.4$ ($\sim15\%$). } & \multirow{3}{6.cm}{HR $\simlt0$ ($\sim45\%$ of the sample) indicates $kT<5$ keV; HR $>0$ ($\sim55\%$) corresponds to $kT>5$ keV.} \\
    &  & \\
    & & \\[2pt]
\hline 
\end{tabular}
\end{center}
\end{table}

\subsection{The origin and composition of the unique source populations in the Galactic Center (NSD)}\label{sec:gc_pops}

As described in Section \ref{sec:hr}, we have identified a distinct concentration of intrinsically hard X-ray point sources in the NSD.  And as shown in 
\ref{sec:shdist}, the density of hard sources within the NSD far exceeds projections based on the analytic SMD model (Figure \ref{fig:sh_dist1}).  These phenomena may well be related, resulting from a unique population of hard X-ray point sources that do not follow the Galactic stellar mass distribution.  

The possibility of such a population of X-ray point sources in the Galactic Center has been previously suggested. \cite{Uchiyama2011} first proposed that the excess diffuse Fe XXV associated with the GCXE could be produced by a new class of sources with strong Fe XXV emission (but in a follow-up study suggested that such a scenario was unlikely, instead favoring a diffuse origin for the hard GCXE \citep{Uchiyama2013}). 
\cite{Heard2013} proposed a dynamically formed binary population in the central $100\, \rm pc \times100\, \rm pc$ as a possible solution for the discrepancy between the SMD and X-ray point source population, a suggestion echoed by \cite{Xu2019} for the NSC.  

What type of source can comprise such a population?  Ongoing follow-up studies are investigating the answer to this question, but below we speculate on several possibilities based on our current findings, supplemented by the results of earlier studies. \\

\subsubsection{Magnetic CVs}\label{subsec:mCVs}

Magnetic CVs, particularly IPs with large WD masses ($M_{\rm WD}$), would fit most of the characteristics of the hard X-ray source population we observe in the central region.  Notably, it can help explain the excess of point sources observed relative to both the SMD and the diffuse Fe XXV emission, as described below.  

\cite{Xu2019} identified a concentration of magnetic CVs with unusually high $M_{\rm WD}$ in the NSC using \chandra\ data, and found that higher $M_{\rm WD}$ is correlated with higher Fe XXV equivalent width (EW$_{6.7}$) -- and lower $<10$ keV X-ray flux, due to their very high shock temperatures.  These results are further supported by \cite{Mondal2024} and \cite{Mondal2025}, who found dozens of magnetic CVs among the \xmm\ Heritage Survey sources with EW$_{6.7}$ much higher than typical values observed in solar-neighborhood sources; their results likewise hint at an anti-correlation between \xmm-measured X-ray flux ($2-10$ keV) and EW$_{6.7}$.  This anti-correlation can help explain the drop in source density that we observe in the NSC region\footnote{As mentioned in Section \ref{sec:shdist} above, $\sim50\%$ of the DR14 catalog sources around $|l^*|<0.15^\circ$
were removed from our sample because they were flagged as potentially spurious or as having unreliable parameters, issues that render those sources unsuitable for our HR analysis.  Should those sources be included, we would indeed observe a sharper peak around the location corresponding to the NSC.  However, within the central $\simlt0.1^{\circ}$ around \sgra, even "bad" sources show a deficit where they should have peaked based on the SMD.  This is a consequence of the limitations of \xmm's angular resolution ($>4\arcsec$ FWHM), which is insufficient to resolve many of the fainter sources in the crowded central parsecs (and our conservative "nearest neighbor" limit for "duplicate" sources).  
} (Figure \ref{fig:badsources}): a large concentration of faint sources would exacerbate the source confusion that we already determined is dominant in that region, leading to a dearth of point sources even as the (seemingly) diffuse Fe XXV emission peaks with the comparatively high EW$_{6.7}$.  This scenario, which is currently being investigated in a follow-up study (Bao et al., in prep), supports a point source origin for the GCXE. 

Past studies of the GCXE have established that: 
\begin{enumerate}
\setlength{\itemsep}{0pt}
    \item The Fe XXV spatial profile shows an excess in the Galactic Center -- compared to the Galactic ridge -- %vis-à-vis 
    with respect to the NIR surface brightness and SMD models \citep{Koyama2007, Uchiyama2011, Nishiyama2013}, 
    \item The measured EW of the diffuse Fe XXV GCXE is $\sim500$ eV -- nearly twice the value observed for magnetic CVs in the solar neighborhood \citep{Nobukawa2016, Xu2016, Koyama2018}, and
    \item The \chandra\ X-ray point source distribution is even more concentrated towards the center compared to the diffuse Fe XXV emission AND the stellar mass \citep{Koyama2007, Hong2009, Muno2009, Heard2013}.
\end{enumerate}
Previously, the differences in the Fe XXV profile with respect to both the stellar distribution and the X-ray point sources were used to argue that the origin of the GCXE is primarily diffuse \citep{Koyama2007, Nobukawa2010, Uchiyama2011, Uchiyama2013, Nishiyama2013}.  \cite{Anastasopoulou2023} showed that the excess Fe XXV in the NSD region can instead be largely accounted for by a population of sources with Fe abundance $\sim1.9\times$ that of the Galactic ridge X-ray sources, a conclusion that is supported by findings of supersolar $Z_{\rm Fe}$ from both NIR and diffuse X-ray metallicity estimates in the Galactic Center \citep{Koyama2007, Uchiyama2013, Feldmeier-Krause2017, Nogueras-Lara2020, Schodel2020, Schultheis2021}.  Our analysis neither refutes nor supports the presence of such a population, as we find that higher Fe abundance ($Z_{\rm Fe}$) does not significantly impact the mean observed HR (cf. Section \ref{sec:hr}).  However, a large concentration of faint IPs with high $M_{\rm WD}$ and elevated $Z_{\rm Fe}$ (and correspondingly large EW$_{6.7}$) in the NSD would simultaneously explain the GCXE Fe XXV excess with respect to the stellar mass distribution, the high EW$_{6.7}$ of the GCXE, and the \chandra\ point source density excess with respect to the diffuse Fe XXV emission.  It is also consistent with the hardness of the X-ray sources we observe in the NSD (which largely overlays the GCXE footprint).

An open question is whether the high stellar density in the NSD can enhance the rate of accreting binaries formed through tidal capture or exchange interactions, potentially creating an excess over the population expected through the canonical common-envelope channel, as suggested by \cite{Heard2013} for the central 100 pc.    
\cite{Muno2009} already noticed an overabundance of point sources along the Arches, Quintuplet, and Galactic Center star clusters, ascribing them to young massive stars.  Our HR analysis rules out a young star contribution to this excess, as the sources we observe in excess of the SMD prediction are much too hard for such a classification.  And unlike the \chandra-based surveys, we also do not observe the hard X-ray source overdensity in these dense star clusters
-- likely due to \xmm's poorer spatial resolution, which hinders source detection in the most crowded central region; we instead observe a dearth of sources at the location of the NSC (Figure \ref{fig:badsources}).  However, an association with those dense star clusters is still possible, perhaps even likely: in crowded cluster environments, dynamical processes, like tidal capture and exchange interactions, may dominate compact binary formation.  Furthermore,  \cite{Uchiyama2011} found an additional Fe XXV excess in the regions $|l|<0.2^{\circ}$ ($\sim$ consistent with the NSC considering the resolution of $0.1^{\circ}$) and $0.1^{\circ}<l<0.3^{\circ}$ (corresponding to the Arches and Quintuplet clusters), hinting at an association between the diffuse GCXE Fe XXV emission and the hard point sources.  
The dynamically-formed-CVs scenario is also consistent with higher-than-average $Z_{\rm Fe}$, which (on its own or in conjunction with the previously presented high-$M_{\rm WD}$ IP scenario) can help explain the overabundance of diffuse Fe XXV in the GCXE: it is plausible that accreting binaries formed through dynamical processes -- such as tidal capture or exchange interactions --  in young star clusters should 
contain 
donors with enhanced $Z_{\rm Fe}$ relative to the typically older stellar populations in other regions of the Galaxy.  However, it is unclear whether the density in the NSD \emph{outside} the aforementioned dense star clusters is high enough to spur dynamical binary formation.  Future theoretical/analytical work may be able to test this hypothesis.

We believe that IPs are better candidates than polars for the hard NSD X-ray population because they have not only higher average thermal bremsstrahlung temperatures ($\langle kT \rangle \approx 34$ keV, vs $\approx11$ keV for polars \citep{Xu2016}), but also additional obscuration of soft X-ray emission.  Unlike polars, IPs form accretion disks that are truncated at the inner edge, from which the accretion flow follows the magnetic field lines, forming "accretion curtains" -- sheets of plasma that can block much of the emission emanating from the accretion column towards the observer line-of-sight.  The soft X-ray emission is far more susceptible to this obscuration, which can therefore make IP spectra appear harder.  There is no equivalent mechanism for polars that would explain such a hard HR distribution.  While the average HRs derived from our thermal model simulations (Figure \ref{fig:simhr}) may not be hard enough to account for the hardest (e.g., HR $>0.4$ at $N_{\rm H}<10^{23}$ cm$^{-2}$) 
of the point sources observed in the NSD, those simulations did not consider the partial covering by the accretion curtains exhibited by many IPs.  When we added a partial covering to otherwise identical models, we found that the HRs were systematically higher.  For example, assuming an \texttt{APEC} spectrum with $kT=36$ keV, $N_{\rm H}=10^{23}$ cm$^{-2}$, and 95\%\footnote{Note that IPs with larger $\dot M$ -- as can be expected for CVs with younger donors -- tend to have shorter accretion column shock height \citep{Salcedo2024, Vermette2023} and are therefore more susceptible to being obscured by the accretion curtains (in addition to having higher bremsstrahlung temperatures due to the reduced cooling).  Large covering fractions ($>90\%$) have been measured for field IPs \citep{Worpel2020} and are even more likely in the NSD, with its substantial population of young stars.} partial covering of $N_{H, \,\rm pc}=5\times10^{23}$ cm$^{-2}$, we derive an HR of $\sim0.82$, consistent with some of the hardest NSD sources; an equivalent model without the covering fraction, and the same $kT$ and $N_{\rm H}$, yielded HR $\sim0.36$.

All things considered, magnetic CVs -- particularly IPs -- may be the most plausible explanation for not only the hard (HR $>0.4$) X-ray sources in the Galactic Center, but also the diffuse GCXE.  
The presence of a large "excess" population of IPs with high $M_{\rm WD}$ (dynamically-formed or otherwise) can negate the need for a diffuse, hot plasma to explain the GCXE origin \citep{Uchiyama2011}; such a hot plasma would not remain gravitationally bound to the Galactic Center \citep{Nishiyama2013}, raising thorny questions about its source.  A point source population is overall more plausible, particularly in light of our findings.

%\smallskip

\subsubsection{Pulsars}

As mentioned in the Introduction, a large population of pulsars, including millisecond pulsars (MSPs), are thought to inhabit the Galactic Center region \citep{Pfahl2004,Ye2022}, deposited there by infalling globular clusters \citep{Brandt2015, Ye2022}.  MSPs have been proposed as the origin of the GeV excess detected by \fermi-LAT in the Galactic Center (with dark matter annihilation a proposed but highly disputed alternative  \citep{Hooper2011, Gordon2013}).  MSPs can be quite hard ($\Gamma<1$ \citep{Sullivan2025}), with corresponding HR exceeding 0.4 in the NSD.  Although we cannot distinguish between MSPs and IPs based on HR alone, our follow-up study using X-ray spectra (Bao et al., in prep) will investigate whether a substantial population of MSPs is present in our sample by evaluating the presence of Fe emission lines (which IPs have but pulsars generally do not).  If MSPs comprise most of our sources, it could indicate that the Fe XXV emission of the GCXE does not originate from unresolved point sources; we consider this scenario unlikely.

\subsubsection{Soft X-ray sources: more CVs?}

As shown in Figure \ref{fig:hr3}, the NSD X-ray source population, despite peaking at HR $>0.4$, also includes a significant fraction of sources with $-0.65<\rm HR < 0.4$.  
For the purposes of this discussion, we define sources in the NSD footprint ($|l|<1.3^{\circ}$, $|b|<0.3^{\circ}$) with HR $<0.4$ as "soft", even though some of the candidate source types might traditionally be viewed as "hard" (e.g., even sources with $kT>10$ keV would fall below the threshold at $N_{\rm H}<8\times10^{22}$ cm$^{-2}$).  
In contrast to the hard sources (HR $>0.4$), the density distribution of the soft sources agrees much better with the SMD.  

DNe, as well as polars, are the most likely dominant contributors to the sizable population with $\rm HR\simlt 0.4$ in the NSD (Figure \ref{fig:hr3}).  While a small fraction of other soft X-ray sources, like symbiotic stars and Wolf-Rayet (WR) colliding-wind binaries, may also be represented among the soft X-ray populations, those source classes are only rarely X-ray bright and hard enough to be detected in our NSD sample, considering how strongly the high absorption ($N_{\rm H}>2\times10^{22}$ cm$^{-2}$) affects the soft X-ray band \citep{Luna2013, Xu2016, Anastasopoulou2024}.  
DNe and polars may be distinguishable in our follow-up study using Fe line diagnostics (Bao et al., in prep).

\subsubsection{LMXBs and HMXBs}

Although much rarer than CVs, LMXBs are likely present in our sample in small numbers.  In particular, those hosting neutron stars (NS-LMXBs) are often bright enough, even in quiescence, to be detected in our survey \citep{Armas-Padilla2014}.  In the inner $\sim10\arcmin$ of the Galaxy, we have been sensitive to X-ray outbursts -- which NS-LMXBs frequently produce -- for the past two decades through the \swift/XRT monitoring program \citep{Mori2021}, but outside that region, all-sky X-ray monitors are not sensitive enough to detect fainter outbursts \citep{Yan2015, Mori2022}.  This makes LMXBs outside the central $\sim10\arcmin$ challenging to identify.  
The large concentration of quiescent BH-LMXBs believed to reside in the central pc ($r<0.5\arcmin$; \cite{Morris1993, Generozov2018, Hailey2018, Mori2021}) is likely too dense and faint to be detected by \xmm.  However, 
thousands of isolated NSs and hundreds of NS-LMXBs are predicted to inhabit the NSC, and indeed NS-LMXBs have been identified out to a distance of $\sim55$ pc ($\sim0.4^{\circ}$) from \sgra\ \citep{Mori2021}.  Altogether, $\sim15\%$ of all identified LMXBs in our Galaxy are found within $\sim0.4^{\circ}$ of \sgra\ \citep{Mori2021, Fortin2024, Mandel2025a}, a significant overconcentration compared to the Galactic stellar mass distribution (although it is not clear whether this excess extends throughout the NSD \citep{Revnivtsev2008}).

If we assume that -- as multiple studies have proposed -- thousands of hitherto-undetected (or unclassified) LMXBs are present in our Galaxy \citep{Kiel2006, Jonker2011, van-Haaften2015, Heinke2024}, and that this population of quiescent, faint LMXBs is similarly concentrated towards the Galactic Center, then it is plausible that our survey may contain up to hundreds of faint LMXBs, particularly of the NS variety.  The latter category includes accreting MSPs \citep{Di-Salvo2022} and ultra-compact X-ray binaries (UCXBs) \citep{Sazonov2020}.  
Quiescent LMXBs typically show soft, non-thermal spectra above 2 keV \citep{Hailey2018}.  Their hardness ratios would likely fall below HR $=0.4$ throughout most of our survey region.  If present, these sources likely comprise a sub-dominant fraction of the soft sources in our survey, far outnumbered by CVs. 

HMXBs are even rarer than their low-mass counterparts.  However, unlike LMXBs, they are expected to trace regions hosting recent star formation activity \citep{Sazonov2020}, which makes the lack of HMXBs in the Galactic Center  \citep{Laycock2005, Mandel2025a} -- with its sizable population of young massive stars -- surprising.  This dearth may be a consequence of the difficulties in identifying HMXBs in the crowded Galactic Center, which is opaque in the optical band and where a definite association with an appropriately bright hot NIR star is challenging to prove \citep{Mandel2025a}; HMXBs do not typically undergo the explosive transient events that are characteristic of LMXBs and often lead to their X-ray detection.  Unlike quiescent LMXBs, HMXBs can be extremely hard (e.g., $\Gamma=0.3$ \citep{Mandel2025a}).  As such, they may be present among our hard NSD sources (HR $>0.4$), but given their rarity, they likely do not represent a significant fraction of those sources.

\subsection{Implications for the X-ray source populations underlying the GRXE}\label{sec:grxe_pops}

A number of studies have attempted to discern the composition of the GRXE, often with conflicting results \citep{Revnivtsev2009, Hong2012, Warwick2014, Xu2016, Schmitt2022}.  However, most of those studies proposed two main classes of sources, in varying ratios:
\begin{itemize}[, itemsep=3pt, topsep=3pt, parsep=0pt]
    \item CVs, particularly polars and DNe: The diffuse GRXE, which is now believed to be comprised of faint unresolved point sources \citep{Revnivtsev2009}, is consistent with a hot, collisionally-ionized, optically thin thermal plasma with measured $kT$ between $\sim5-15$ keV \citep{Tanaka2002, Yuasa2012, Uchiyama2013, Nobukawa2016}.  The spectral properties of the diffuse emission are broadly consistent with both polars and DNe.  Several studies consider them the dominant contributors to the GRXE (e.g., \cite{Hong2012} and \cite{Schmitt2022}), although there is also disagreement over whether DNe or magnetic CVs are predominant \citep{Hong2012, Xu2016}.  
    \item Active binaries (ABs): ABs were proposed as a major contributor to the GRXE by \cite{Worrall1983}, but this has turned out to be controversial: while \cite{Revnivtsev2006} and \cite{Warwick2014} support ABs as significant or even dominant contributors to the GRXE, \cite{Hong2012} and \cite{Schmitt2022} strongly disagree.  ABs are generally fainter and much softer than CVs, with $kT\sim$ a few keV \citep{Xu2016}.  \cite{Warwick2014} suggests that they can still contribute to the hard GRXE during flaring episodes; they may compensate for their comparatively lower X-ray luminosity by their much greater numbers (although this does not bode well for their detection in our point source survey).
    
\end{itemize}

We find that the largest concentration of sources outside the NSD ($\sim40\%$) have HR between $-0.4$ and 0.25 (Figure \ref{fig:hr3}), consistent with the HR for thermal sources with $kT>4$ keV at $N_{\rm H}\sim3-5\times10^{22}$ cm$^{-2}$.  These results indicate that DNe and magnetic CVs are favored as the dominant X-ray source population in the Galactic disk, outside the NSD.  The presence of sources with HR $>0.2$ ($40\%$ of the sample) may require a contribution from 
IPs (with their characteristic partial covering), since the unobscured thermal models do not yield HR $>0.2$ where $N_{\rm H}$ is below $\sim5\times10^{22}$ cm$^{-2}$.  While this study cannot effectively distinguish between DNe and polars, our follow-up study using spectroscopy may be able to determine the fractional contribution for each class.

Another prominent contributor may be symbiotic stars (SSs).  \cite{Luna2013} and \cite{Xu2016} previously suggested a possible minor SS component to the GRXE emission, but \cite{Bao2026} showed that up to 20\% of the point sources in the \xmm\ Heritage Survey that were meaningfully associated with \gaia\ counterparts may be SSs.  Those sources were found to have a typical temperature of $kT\sim5-6$ keV and a large Fe XXV EW ($800-900$ eV).  It is as yet unclear whether the prominence of this SS population (or indeed their spectral characteristics) holds for the extended Galactic disk, as the \gaia-matched sample is severely distance-limited; however, even foreground sources with the spectral parameters measured by \cite{Bao2026} would likely remain in our sample, as HR$_{\rm fgd}$ cannot effectively distinguish between hard foreground sources and soft Galactic Center sources.  
Upcoming NIR surveys by the Nancy Grace Roman Space Telescope (\textit{Roman}) may allow us to determine their prevalence.

Among the sources in our sample with HR $<-0.3$ ($\sim27\%$), we expect some contribution from ABs; at the high end of their luminosity range, they should be detectable in our sample.  Likewise, a small contribution from young massive stars (including O/B and WR stars \citep{Anastasopoulou2024}) may be present.  Finally, AGN may comprise a minor component, though given the high levels of absorption present not only in our foreground but also \emph{behind} our line-of-sight towards the Galactic Center, their contribution is likely negligible.   \\

\section{Conclusions} \label{sec:conclude}

We present a comprehensive analysis of the X-ray point sources detected by \xmm\ within the central $20^{\circ}\times2^{\circ}$ of the Galactic plane.  We summarize our findings below:

\begin{itemize}
    \item There is a concentration of hard X-ray sources within the footprint of the NSD ($|l|<1.3^{\circ}$ and $|b|<0.3^{\circ}$).  The hardness appears to be at least in part intrinsic, as the high $N_{\rm H}$ alone cannot account for it.  This indicates that the X-ray population inside the NSD is distinct from that of the Galactic plane.
    \item This concentration of hard sources does not follow the SMD; it is more dense within the NSD footprint in general and especially within $\sim0.2^{\circ}$ ($25$ pc) of \sgra, hinting at possible dynamical formation channels in the dense young star clusters around \sgra.
    \item The concentration of hard sources in the NSD most likely consists of IPs with high $M_{\rm WD}$, possibly with a contribution from MSPs.  
    \item The distribution of softer X-ray sources (HR$<0.4$ for PN and HR$<0.25$ for MOS) is largely consistent with the SMD.   
    \item Outside the NSD, the X-ray population appears to be consistent primarily with thermal sources of mean temperature $kT\sim10$ keV.  This indicates a prevalence of 
    polars and DNe, as found by previous studies, with a contribution by IPs likely to account for the harder sources (and possibly SSs for the softer sources).
\end{itemize}

\begin{acknowledgments}
This research has made use of data obtained from the 4XMM XMM-Newton Serendipitous Source Catalog compiled by the 10 institutes of the XMM-Newton Survey Science Centre selected by ESA. 
We would like to express our gratitude to Ada Nebot for helpful discussions about the catalog features.  
The authors thank Konstantina Anastasopoulou and Mattia Sormani for providing the SMD model used in our analysis.  
This material is based upon work supported by the National Science Foundation Graduate Research Fellowship under Grant No. DGE 2036197. 
SM gratefully acknowledges support from the Columbia University Provost Diversity Fellows Program. 
GP and SM acknowledge financial support from the European Research Council (ERC) under the European Union’s Horizon 2020 research and innovation program HotMilk (grant agreement No. 865637). GP and SM also acknowledge support from Bando per il Finanziamento della Ricerca Fondamentale 2022 dell’Istituto Nazionale di Astrofisica (INAF): GO Large program and from the Framework per l’Attrazione e il Rafforzamento delle Eccellenze (FARE) per la ricerca in Italia (R20L5S39T9).  SM also acknowledges support from grant NASA ADAP 80NSSC24K0666. MM acknowledges support from NASA ADAP grant 80NSSC24K0639 to UCLA.  
SM acknowledges support from grant NASA ADAP 80NSSC24K0666 and NASA NuSTAR data analysis funding 1729326.

\end{acknowledgments}


\vspace{5mm}
\facilities{XMM(MOS1, MOS2, and EPIC-PN), }


\software{astropy \citep{2013A&A...558A..33A,2018AJ....156..123A},  
        %XSPEC
%          Cloudy \citep{2013RMxAA..49..137F}, 
%          Source Extractor \citep{1996A&AS..117..393B}
          }

\bibliography{main}{}
\bibliographystyle{aasjournal}

%\clearpage

\appendix

\section{Source selection and spatial distribution of rejected sources}\label{sec:AA}

We began by selecting all sources within the DR14 catalog that fall within the \xmm\ Heritage Survey region, i.e. $|l|\simlt10^{\circ}$, $|b|<1^{\circ}$.  Without applying any additional filters, those catalog entries numbered 48073.  
We identified three main sources of stray light background that affect the \xmm\ Heritage Survey data (green circles in Figure \ref{fig:exp}).  We found that most of the spurious sources that constitute the characteristic arcs generated by stray light are flagged as "located in a area where spurious detection may occur" (\textsc{SUM\_FLAG}=3).  We therefore removed sources that are flagged as either "possibly spurious" (\textsc{SUM\_FLAG}=2), "located in a area where spurious detection may occur" (\textsc{SUM\_FLAG}=3), or both (\textsc{SUM\_FLAG}=4).  To maximize the reliability of our analysis, we further excluded sources flagged "source parameters may be affected" (\textsc{SUM\_FLAG}=1), retaining only those sources whose \textsc{SUM\_FLAG}=0 ("good sources").  After also removing all sources flagged as "extended" or "confused", we were left with a total of 25711 "clean" point sources; this tally included many sources that were detected in multiple observations, and hence listed in multiple entries.

To remove "duplicate" entries of the same source from different observations, we systematically reviewed every catalog source that withstood the purges described above and was also marked as having multiple detections (\textsc{N\_DETECTIONS}$>1$).  We collated the catalog entries that shared the same source ID (\textsc{SRCID}), and selected the entry with the longest exposure (EPIC PN, MOS1, and MOS2 combined) to represent the source in our analysis; other entries with the same source ID were removed from our sample, although their data were still added to the retained "representative" entry if they met the criteria described below.  We also removed nine additional sources whose distance to their nearest neighbor was small enough ($<6\arcsec$) to suggest a duplicate entry, despite having a distinct source ID.

For each unique source, we collated the counts, count rates and exposures from every observation that yielded at least 10 counts and/or lasted at least 10 ks.  These thresholds were selected based on the completeness limit of our dataset ($\sim10^{-3}$ cts s$^{-1}$; Figure \ref{fig:ctrt_all}).  Final count rates utilized in our analysis were calculated using the exposure-weighted average from all observations that met the above criteria.  We treated the PN and MOS data separately, given their diverging effective areas; however, MOS1 and MOS2 data were combined (where both were available). 

\subsection{Foreground source identification and distribution}\label{subsec:AAa} 

As we aim to analyze here the populations of X-ray sources in and near the Galactic Center, we want to remove foreground sources from our sample, particularly those in the Solar neighborhood which can bias the observed distributions.  To that end, we defined a "foreground" hardness ratio HR$_{\rm fgd}$, with a soft band of $0.5-1$ keV and a hard band of $2-12$ keV.  High column density $N_{\rm H}$ toward the Galactic Center results in very high absorption at energies $<1$ keV for all except foreground sources, making HR$_{\rm fgd}$ an effective way to distinguish them.  For a thermal (\texttt{APEC}) spectrum with $kT=1$ keV (consistent with e.g. active stars, ABs, YSOs, etc.; \cite{Bao2025}) and $N_{\rm H}=10^{22}$ cm$^{-2}$ (as expected for sources out to 3.5 kpc according to \cite{Pohl2022}), HR$_{\rm fgd}=-0.1$, which we adopt as our foreground threshold for PN. For MOS, that threshold increases to HR$_{\rm fgd}=0$ after accounting for differences in effective area.  These thresholds are conservative and some foreground sources -- particularly harder ones -- may remain in the sample after they are applied. 

\begin{figure}
    \centering
    \includegraphics[width=0.75\linewidth]{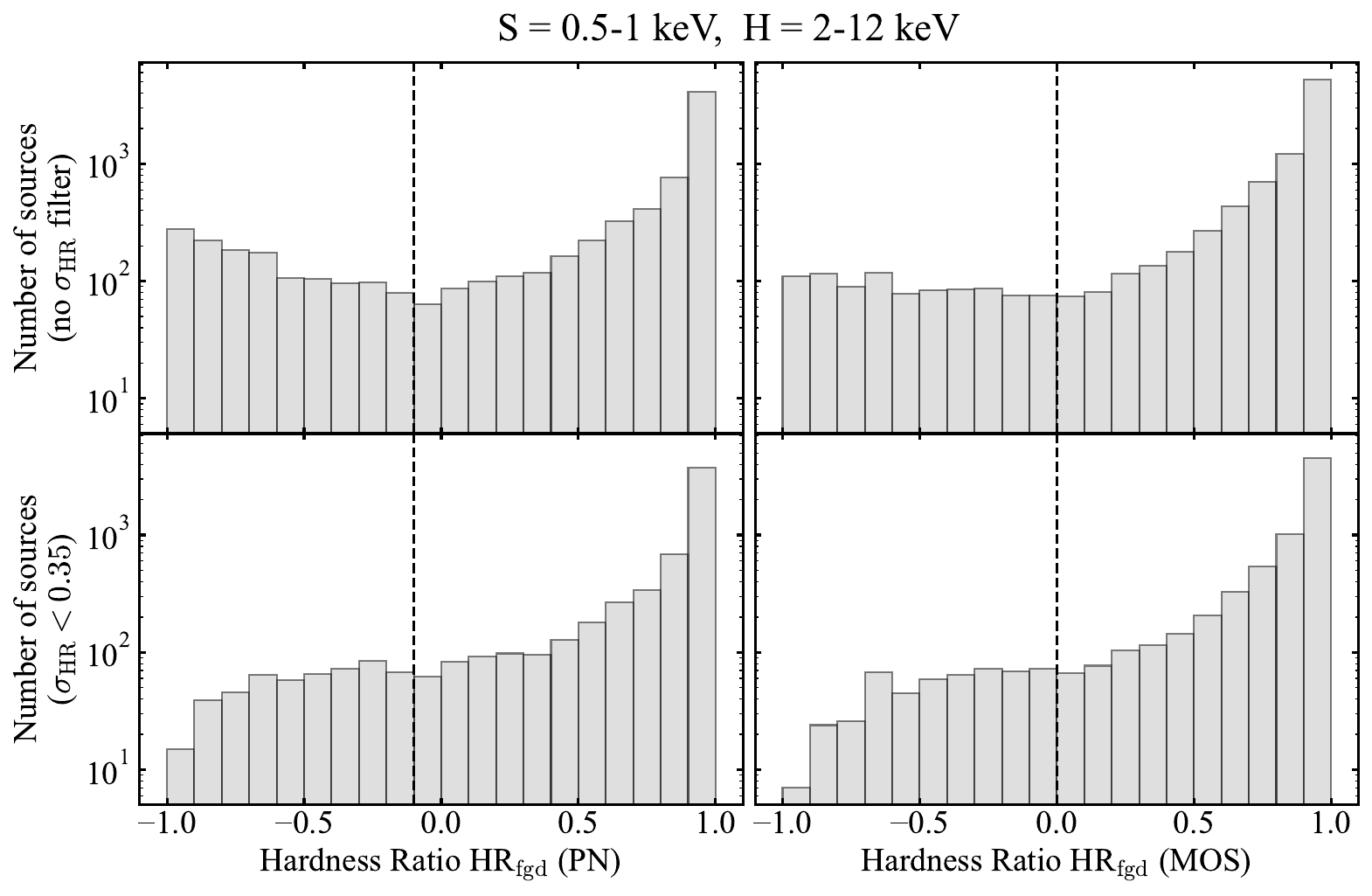}
    \caption{"Foreground" hardness ratio HR$_{\rm fgd}$ distributions for PN (left) and MOS (right), for all "clean" sources with $\sigma_{\rm HR_{fgd}}<0.35$ (top) and with the additional stipulation of $\sigma_{\rm HR}<0.35$ (bottom).  The majority of sources with HR$_{\rm fgd}$ below the foreground threshold (dashed lines; -0.1 for PN and 0 for MOS) are excluded from the sample with $\sigma_{\rm HR}<0.35$.  Remaining sources below the %foreground 
    threshold (i.e., to the left of the dashed line) are considered "foreground" and removed from the final sample analyzed in this work.}
    \label{fig:hr6}
\end{figure}

A significant fraction of the sources in the \xmm\ Galactic Plane Heritage Survey of the central $20^{\circ}\times2^{\circ}$ have HR$_{\rm fgd}$ consistent with foreground sources 
(Figure \ref{fig:hr6}, top).  However, the majority of those sources are already excluded from our sample because they have HR uncertainty  $\texorpdfstring{\sigma\textsubscript{HR}}{}>0.35$ (Figure \ref{fig:hr6}, bottom), generally due to low counts in the $2-12$ keV band.  Considering only sources in our "clean" sample with $\texorpdfstring{\sigma\textsubscript{HR}}{}<0.35$, apparent foreground sources\footnote{We consider as "foreground" sources those with HR$_{\rm fgd}<-0.1$ (0) for PN (MOS) AND $\sigma_{\rm HR_{fgd}}>0.35$; where $\sigma_{\rm HR_{fgd}}$ (the uncertainty in HR$_{\rm fgd}$) is large, HR$_{\rm fgd}$ is no longer a reliable identifier of foreground sources. } comprise only 5\% (110 sources) of the population in the NSD, and 7\% (397) of the sources outside the NSD (MOS).  Removing those sources does not significantly impact the mean or median values of the HR distributions inside vs outside the NSD.  For example, before excluding likely foreground sources, the corresponding median HR was 0.17 inside the NSD and -0.18 outside; after removing the foreground sources, median HR changed to 0.19 in the NSD and -0.15 outside the NSD.

One notable result of our foreground analysis is the concentration of apparent foreground sources ($\sim17\%$ of the "clean" sample) detected at the location of NGC 6405 (the "Butterfly Cluster"), near $l\sim-3.5$ (Figure \ref{fig:badsources}).  Our detection of this cluster suggests that our HR$_{\rm fgd}$ filter is indeed reliable; NGC 6405 is a young open cluster located at a distance of only 
$\sim461$ pc \citep{Gao2018}.

\subsection{Spatial distribution of excluded sources and the density profile}\label{subsec:AAb}

\begin{figure}
    \centering
    \includegraphics[width=0.9\linewidth]{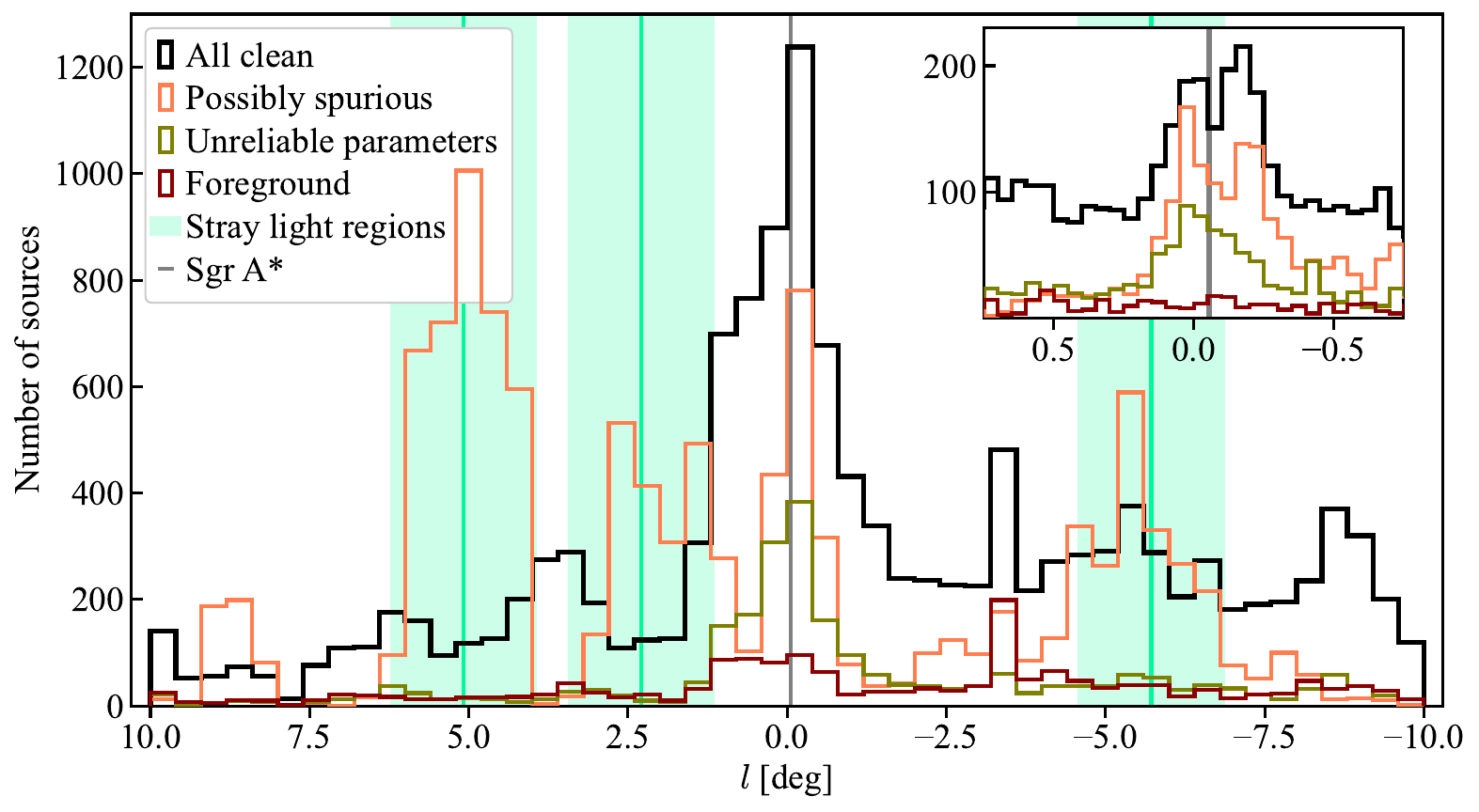}
    \caption{Distributions of "clean" sources (black), potentially spurious sources (orange), sources with unreliable parameters (olive), and likely foreground sources (dark red) over $l$.   
    The inset on the top right shows the distribution around $l=0$ in greater detail.  All source groups show a deficit around \sgra\ (grey line), except foreground sources which we expect to be mostly uniformly distributed (note the NGC 6405 foreground cluster at $l\sim-3.5$).  "Possibly spurious" sources are concentrated in regions with stray light background contamination (green columns), but also in the central $\sim0.15^{\circ}$ ($\sim20$ pc) around \sgra.  Sources with unreliable parameters are concentrated only within $l\sim0.15^{\circ}$ around \sgra.  
    }
    \label{fig:badsources}
\end{figure}

As described above, we removed from our analysis all sources that are flagged as potentially spurious or as having unreliable parameters.  The former group is primarily concentrated in regions affected by stray light background, but also shows a peak at $|l|\simlt0.15^{\circ}$ ($\sim20$ pc); the latter is concentrated only around $|l|\simlt0.15^{\circ}$, a region consistent with the NSC (Figure \ref{fig:badsources}).  The high background and source crowding in the NSC make it difficult to derive reliable parameters even for robustly detected sources.  However, removing sources with unreliable parameters (and by the same token, sources marked \emph{potentially} spurious that are in fact \emph{not} spurious) from our sample also hinders our ability to constrain the scale heights of our source distributions (Section \ref{sec:shdist}), given that the removed sources are unevenly distributed.  Should we add the sources with unreliable parameters to our spatial distribution, the scale heights would be smaller than what we currently measure due to the increased concentration within $|l|\simlt0.15^{\circ}$, which falls below our currently measured scale heights ($h_{l,\rm s}=2^{\circ}$ [$\sim280$ pc], $h_{l,\rm h}=0.66^{\circ}$ [$\sim90$ pc] for the hard and soft source distributions, respectively).  Similarly, removing potential duplicate entries (i.e., all those with a nearest neighbor $<6\arcsec$ offset) is bound to reduce the fraction of retained sources in the most crowded central region, where the separation between sources can often be $<6\arcsec$. 

While all three distributions ("clean" sources, potentially spurious sources, and those with unreliable parameters) are highest near $|l|\sim0.15^{\circ}$, the "peak" is remarkably flat within that region, except for a dip in the number of sources around $0.05^{\circ}-0.1^{\circ}$ from \sgra\ (Figure \ref{fig:badsources}, inset).  Based on the stellar mass distribution, the source density would be expected to peak there; instead, it appears that \xmm's angular resolution reaches its confusion limit within that region, resulting in a drop in the number density of detected sources.

\section{Correction factor for MOS HR}\label{sec:AAa}

\begin{figure}
    \centering
    \includegraphics[width=0.7\linewidth]{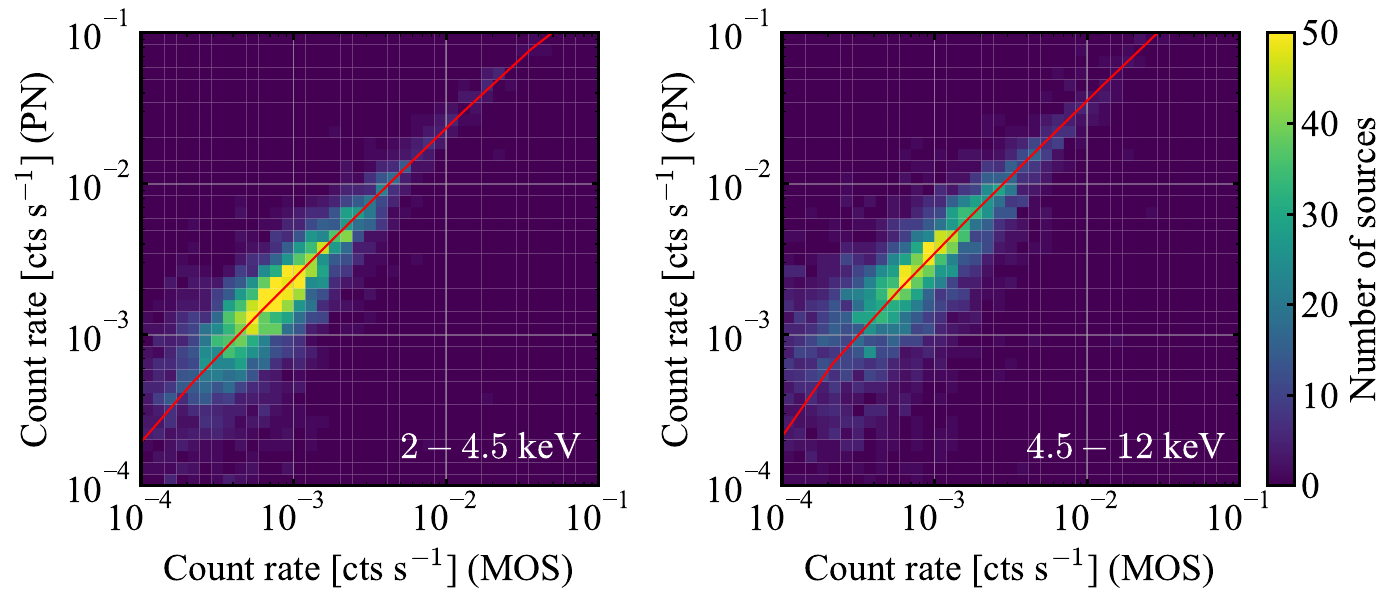}
    \caption{2D density plots showing PN vs MOS count rates in the soft ($2-4.5$ keV; left) and hard ($4.5-12$ keV; right) bands.  The best-fit second-order polynomial is outlined in red.}
    \label{fig:pnvmos_ctrt}
\end{figure}

\begin{figure}
    \centering
    \includegraphics[width=0.7\linewidth]{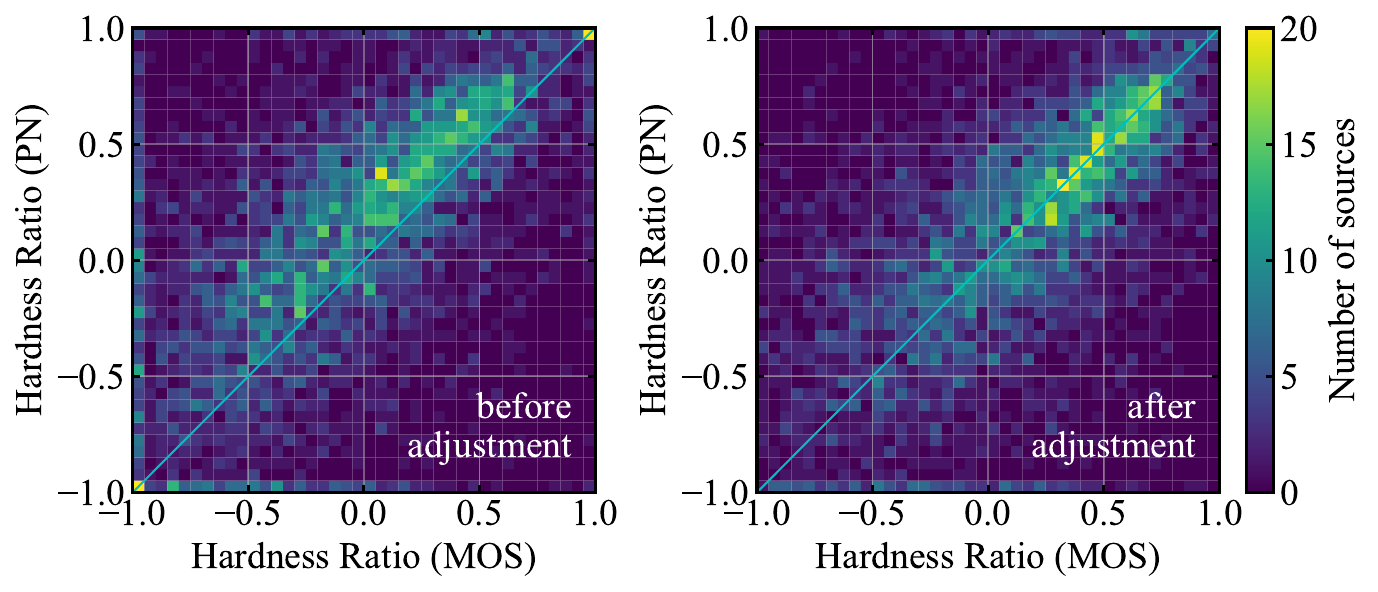}
    \caption{2D density plots showing the hardness ratios HR derived from the PN and MOS data, without modification (left) and after the MOS count rates were scaled using the best-fit functions shown in Figure \ref{fig:pnvmos_ctrt} (right).  The cyan line highlights where HR would be identical for PN and MOS.  After adjusting, the PN and MOS HR is clustered around equivalent values (right).}
    \label{fig:pnvmos_hr}
\end{figure}

For the analysis presented in Sections \ref{subsec:trans} and \ref{subsec:nhsims}, we require HR for the PN and MOS datasets to be equivalent.  This necessitated an adjustment to account for the differences in effective area between the PN and MOS instruments.  The ratio of effective areas for the hard ($4.5-12$ keV) band vs the soft ($2-4.5$ keV) band is substantially higher for PN compared to MOS, leading to systematically higher hard count rates (Figure \ref{fig:pnvmos_ctrt}) -- and consequently, higher HR (Figure \ref{fig:pnvmos_hr}, left) -- for the PN data.

To account for these differences, we determined the best-fit function for the PN vs MOS count rates in the hard and soft bands, respectively.  We found that a second-order polynomial fit most of the data well.  We then used the best-fit functions to modify the MOS count rates and re-calculated the associated HR (Figure \ref{fig:pnvmos_hr}, right).  As expected, the HR values for PN and MOS largely converged.  For a small fraction ($\sim10\%$) of the sample, the adjustment shifted the MOS HR outside the range of [-1,1].  Those sources were excluded from the analysis in Sections \ref{subsec:trans} and \ref{subsec:nhsims}.
\\

\section{Count rate and exposure distributions of \XMM\ catalog sources}\label{sec:AB}

The \xmm\ Heritage Survey was conducted with average exposures of $\sim20-40$ ks per pointing.  However, in certain regions (particularly the central $\sim2^{\circ}$), the available exposures (including from previous archived observations) are much deeper.  Complicating matters, both the diffuse X-ray background and source density are much higher in the same region.  Modeling the competing effects of longer exposures (which improve detection of faint sources) vs high background and source confusion (which hamper detection and/or the extraction of reliable parameters) is non-trivial.  We attempt to correct for the biases by introducing filters in exposure and count rate to retain a more uniform sample.  

\begin{figure}%[h!]
\centering
\includegraphics[width=0.8\textwidth]{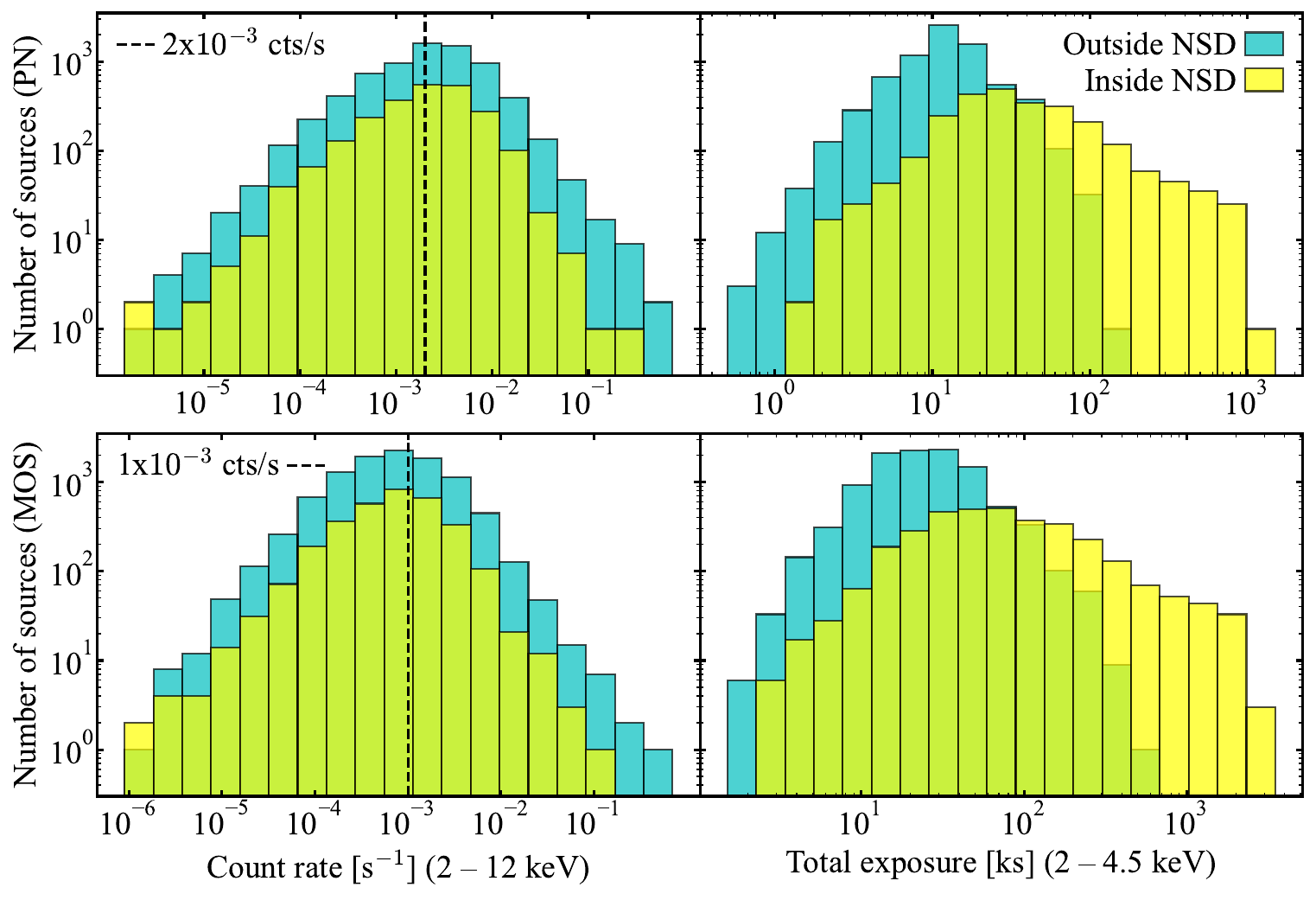}
\caption{Left: Distributions of $2-12$ keV exposure-weighted mean count rates for point sources outside the NSD (cyan) and within $|l|<1.3^{\circ}$ and $|b|<0.3^{\circ}$ (yellow), after filters described in Section \ref{sec:obs} were applied.  The dashed line marks the lower limit for the sample used to test HR robustness (Section \ref{subsec:test_hr}).  Right: Distributions of exposures for all point sources in the \xmm\ survey outside the NSD (cyan) and within $|l|<1.3^{\circ}$ and $|b|<0.3^{\circ}$ (yellow), after filters described in Section \ref{sec:obs} were applied.  PN data are in the top panels; MOS data in the lower panels.    
See Section \ref{subsec:test_hr} for more.
}
\label{fig:ctrt_all}
\end{figure}

\begin{figure}%[h!]
\centering
\includegraphics[width=0.8\textwidth]{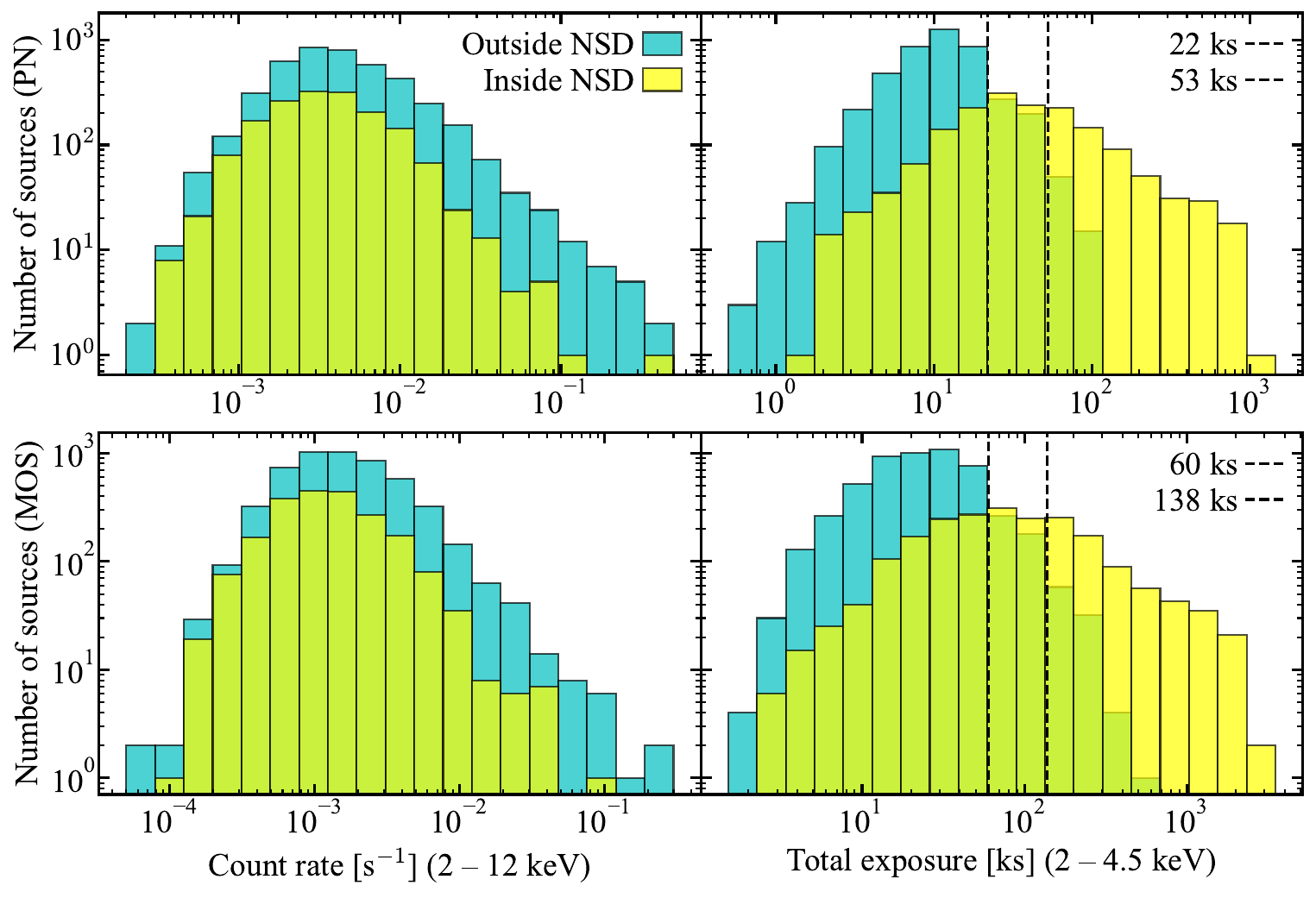}
\caption{Same as Figure \ref{fig:ctrt_all}, except that sources with large uncertainties in hardness ratio ($\sigma\textsubscript{HR}>0.35$), as well as foreground sources, were also removed from this sample (as described in Section \ref{sec:hr}), cutting off many of the fainter sources. 
The dashed lines bracket the exposure range utilized to test the HR distributions in 
Section \ref{subsec:test_hr}.  
}
\label{fig:ctrt_dist}
\end{figure}

Figure \ref{fig:ctrt_all} (left) shows the distribution of count rates for all "clean" sources in our survey, inside (yellow) and outside (cyan) the NSD.  
For the count rate filter introduced in Section \ref{subsec:test_hr}, we applied a minimum threshold of $2\times10^{-3}$ cts s$^{-1}$ for PN data and $1\times10^{-3}$ cts s$^{-1}$ for MOS data (dashed black lines), below which the survey appears incomplete.  This threshold was chosen in order to remove the faintest sources (which are more prone to large statistical uncertainties; cf. Figure \ref{fig:ctrt_dist}) while retaining a sufficient sample for robust statistical comparison between the regions inside and outside the NSD.

The distribution of exposures for all "clean" sources inside (yellow) and outside (cyan) the NSD is shown in Figure \ref{fig:ctrt_all} (right).  
In Figure \ref{fig:ctrt_dist} (right), we additionally exclude sources with HR uncertainty $\texorpdfstring{\sigma\textsubscript{HR}}{}>0.35$.  
We select the range of exposures between the dashed black lines ($22-53$ ks for PN and $60-138$ ks for MOS) to produce a sample in which the sources inside and outside the NSD are approximately uniform in exposure, as the shallowest exposures (which would bias the distributions towards the brightest sources, predominantly outside the NSD) and the longest exposures (which could introduce many more faint sources inside the NSD) are excluded.  This uniform sample was utilized for the second test described in Section \ref{subsec:test_hr}.

\section{Possible effects of observational biases on the log$N$--log$S$ distributions}\label{sec:AC}

As shown in Section \ref{sec:comp}, the best-fit slope $\alpha$ for the log$N$-log$S$ distributions inside and outside the NSD is affected by 
\begin{itemize}[noitemsep, topsep=0pt]
    \item the filters used to select the sample, and 
    \item the choice of flux range used to fit the line.
\end{itemize}
Here we elaborate on these effects.  As in Section \ref{sec:shdist}, we divide the sources into hard and soft groups (with HR above and below $0.4$ for PN and $0.25$ for MOS, respectively), and further divide each group into sources inside ($|l|\simlt1.3^{\circ}$ and $|b|\simlt0.3^{\circ}$) and outside the NSD.

\begin{figure}
    \centering
    \includegraphics[width=0.9\linewidth]{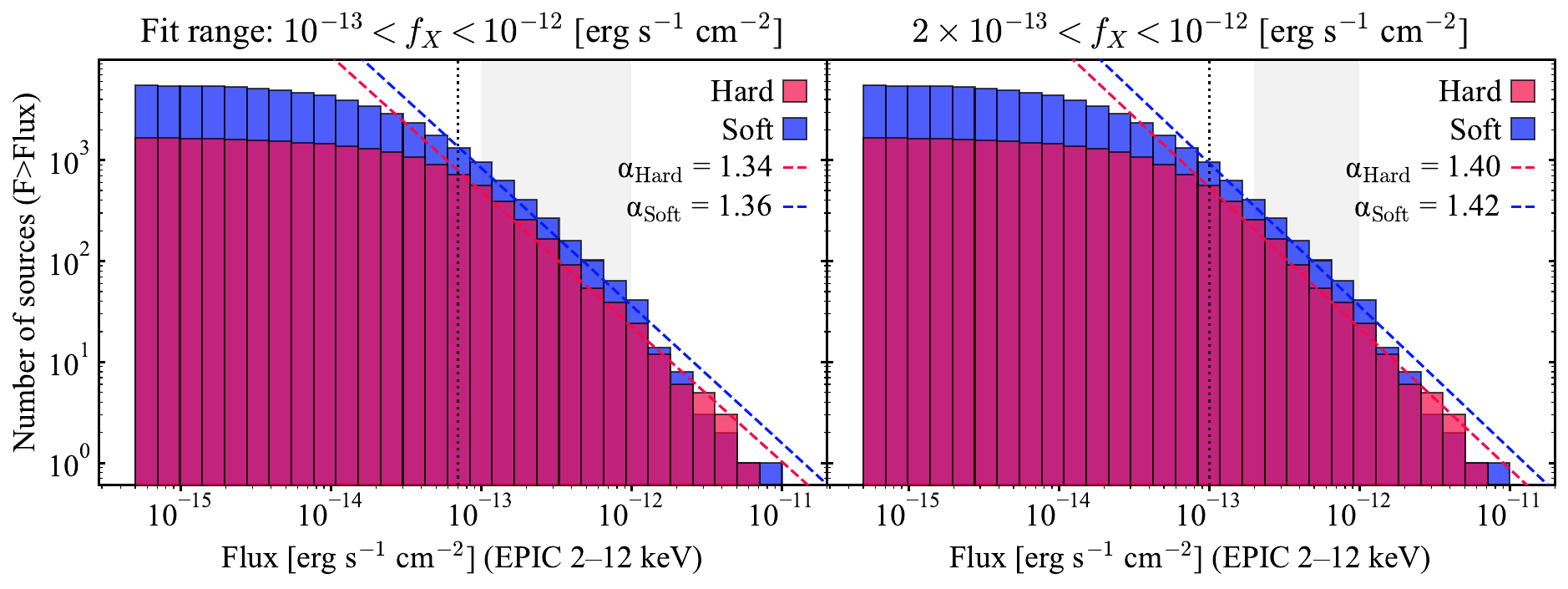}
    \caption{log$N$-log$S$ distribution of hard (red) and soft (blue) sources outside the NSD, with the best-fit slopes ($\alpha$) in the range $10^{-13}<f_X<10^{-12}$ \fluxcgs\ (left) and $2\times10^{-13}<f_X<10^{-12}$ \fluxcgs\ (right).}
    \label{fig:lognlogsout_hardsoft}
\end{figure}

\begin{figure}
    \centering
    \includegraphics[width=0.9\linewidth]{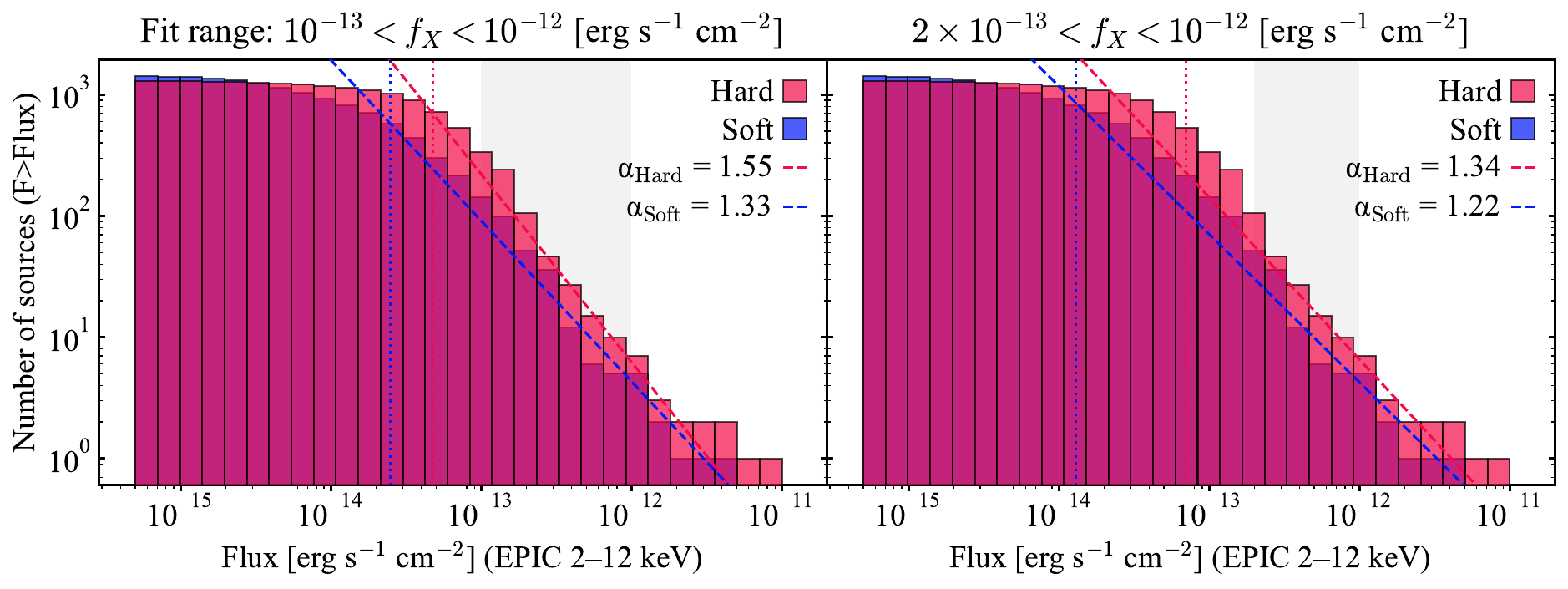}
    \caption{log$N$-log$S$ distribution of hard (red) and soft (blue) sources within the NSD region ($|l|\simlt1.3^{\circ}$ and $|b|\simlt0.3^{\circ}$), with the best-fit slopes ($\alpha$) in the range $10^{-13}<f_X<10^{-12}$ \fluxcgs\ (left) and $2\times10^{-13}<f_X<10^{-12}$ \fluxcgs\ (right).}
    \label{fig:lognlogsnsd_hardsoft}
\end{figure}

As Figures \ref{fig:lognlogsout_hardsoft} and \ref{fig:lognlogsnsd_hardsoft} illustrate, the hard and soft distributions inside and outside the NSD are affected differently when the range over which the log$N$-log$S$ slope ($\alpha$) is fit changes.  Outside the NSD, $\alpha_{\rm soft}$ does not significantly change when the fitting range is expanded(Figure \ref{fig:lognlogsout_hardsoft}).  This is to be expected; the assumed column density ($N_{\rm H}=3\times10^{20}$ cm$^{-2}$) used for the flux extrapolation is closer to the actual values observed outside the NSD.  
However, inside the NSD, the hard distribution 
yields a steeper $\alpha_{\rm hard}$ for the wider fitting range (Figure \ref{fig:lognlogsnsd_hardsoft}, left).  $\alpha_{\rm soft}$ is more consistent with the values derived for the distribution outside the NSD, but only for the wider fitting range; when the range is narrowed, it, too, yields a shallower slope.  These results demonstrate the challenges of extracting statistically robust flux distributions from catalog data where uniform model fitting is employed.  Our future study will incorporate spectral analysis of the point sources in this survey, allowing us to generate more accurate fluxes and evaluate their distributions.

\section{HR distributions for simulated spectra}\label{sec:AE}

\begin{figure}
    \centering
    \includegraphics[trim=0.25cm 0.5cm 0.25cm 0.25cm, width=0.95\linewidth]{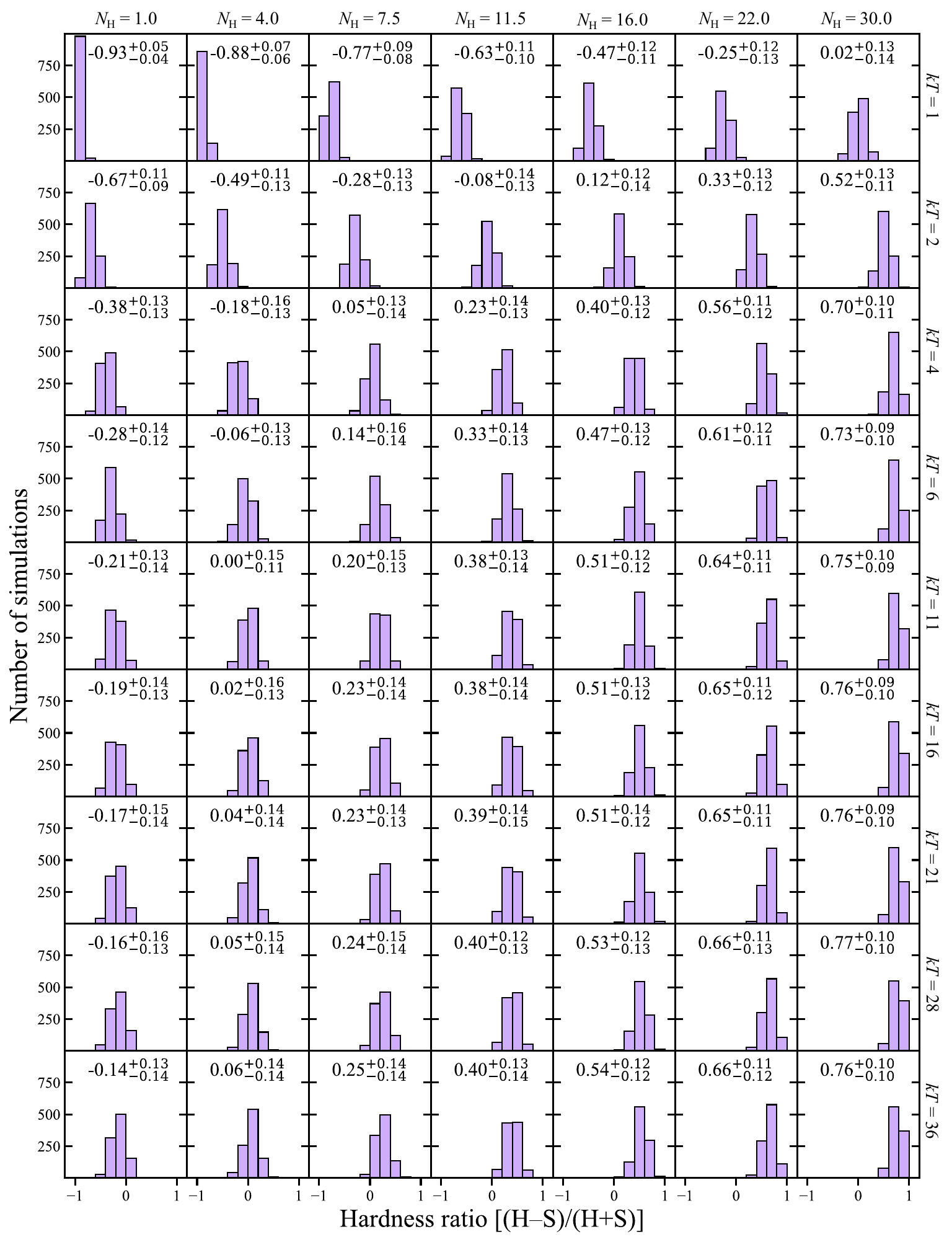}
    \caption{Histograms showing the simulated HR distributions for each \texttt{APEC} spectral model.  The median HR is printed for each distribution along with the $\pm34.1\%$ uncertainty range. $N_{\rm H}$ is listed for each column in units of $10^{22}$ cm$^{-2}$; $kT$ is given for each row in units of keV.}
    \label{fig:hrhist_apec}
\end{figure}

\begin{figure}
    \centering
    \includegraphics[trim=0.25cm 0.25cm 0.25cm 0.25cm, width=0.95\linewidth]{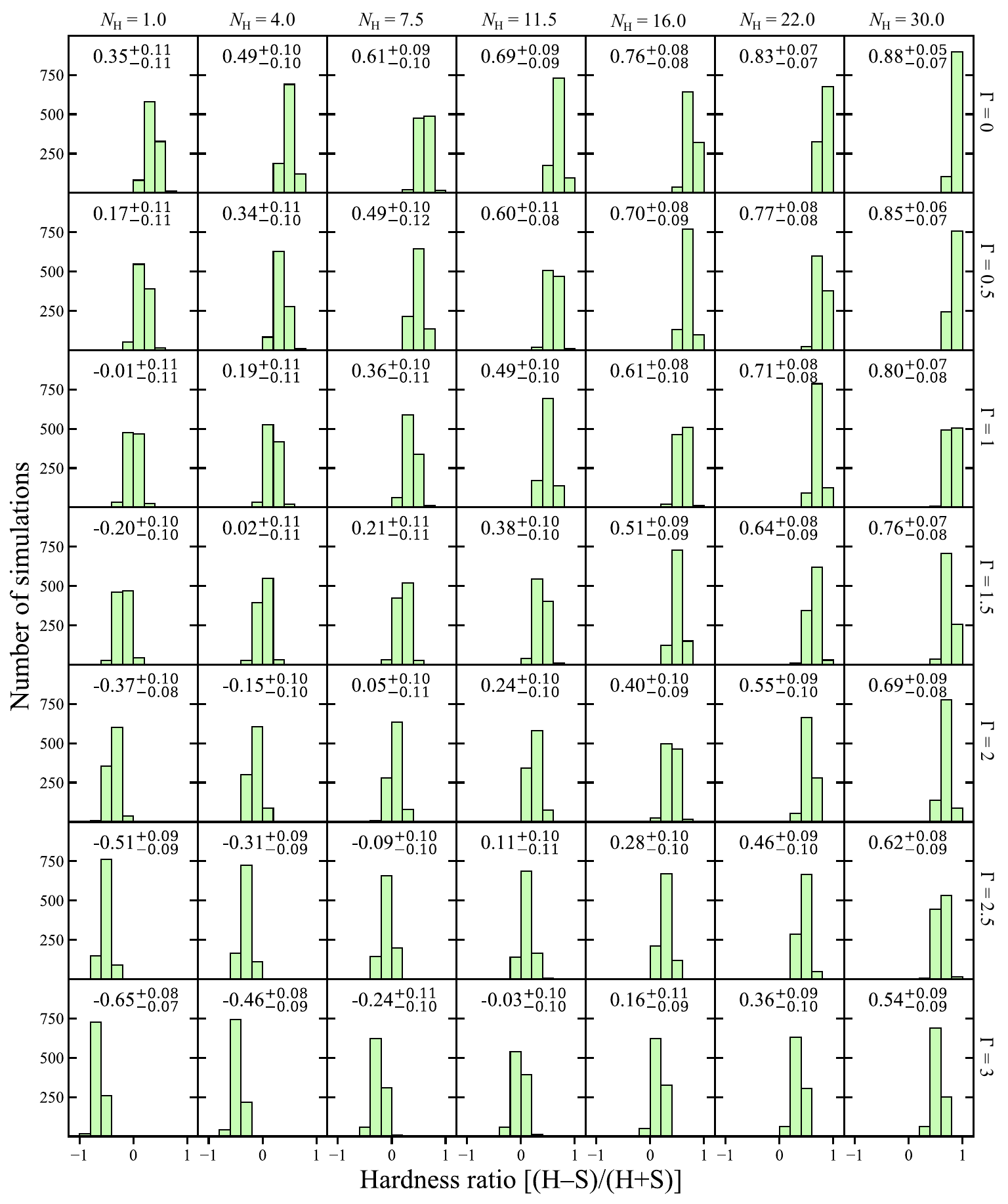}
    \caption{Histograms showing the simulated HR distributions for each \texttt{POWERLAW} spectral model.  The median HR is printed for each distribution along with the $\pm34.1\%$ uncertainty range. $N_{\rm H}$ is listed for each column in units of $10^{22}$ cm$^{-2}$; $\Gamma$ is given for each row.}
    \label{fig:hrhist_pow}
\end{figure}

\end{document}